\documentstyle[preprint,aps]{revtex}
\begin{document}
\input{psfig}
\draft

\title{Antiproton Production and Antideuteron Production Limits 
in Relativistic Heavy Ion Collisions} 

\maketitle
\begin{center}
\bigskip
\mbox{T.A. Armstrong                \unskip,$^{(7,\ast)}$}
\mbox{K.N. Barish                   \unskip,$^{(3)}$}
\mbox{S. Batsoulli                  \unskip,$^{(12)}$}
\mbox{S.J. Bennett                  \unskip,$^{(11)}$}
\mbox{A. Chikanian                  \unskip,$^{(12)}$}
\mbox{S.D. Coe                      \unskip,$^{(12,\dag)}$}
\mbox{T.M. Cormier                  \unskip,$^{(11)}$}
\mbox{R. Davies                     \unskip,$^{(8,\ddag)}$}
\mbox{C.B. Dover                    \unskip,$^{(1,\S)}$}
\mbox{P. Fachini                    \unskip,$^{(11)}$}
\mbox{B. Fadem                      \unskip,$^{(4)}$}
\mbox{L.E. Finch                    \unskip,$^{(12)}$}
\mbox{N.K. George                   \unskip,$^{(12)}$}
\mbox{S.V. Greene                   \unskip,$^{(10)}$}
\mbox{P. Haridas                    \unskip,$^{(6)}$}
\mbox{J.C. Hill                     \unskip,$^{(4)}$}
\mbox{A.S. Hirsch                   \unskip,$^{(8)}$}
\mbox{R. Hoversten                  \unskip,$^{(4)}$}
\mbox{H.Z. Huang                    \unskip,$^{(2)}$}
\mbox{B.S. Kumar                    \unskip,$^{(12,\|)}$}
\mbox{T. Lainis			    \unskip,$^{(9)}$}
\mbox{J.G. Lajoie                   \unskip,$^{(4)}$}
\mbox{Q. Li                         \unskip,$^{(11)}$}
\mbox{B. Libby                      \unskip,$^{(4,\P\P)}$}
\mbox{R.D. Majka                    \unskip,$^{(12)}$}
\mbox{T.E. Miller                   \unskip,$^{(10)}$}
\mbox{M.G. Munhoz                   \unskip,$^{(11)}$}
\mbox{J.L. Nagle                    \unskip,$^{(12,\ast\ast)}$}
\mbox{I.A. Pless                    \unskip,$^{(6)}$}
\mbox{J.K. Pope                     \unskip,$^{(12,\dag\dag)}$}
\mbox{N.T. Porile                   \unskip,$^{(8)}$}
\mbox{C.A. Pruneau                  \unskip,$^{(11)}$}
\mbox{M.S.Z. Rabin                  \unskip,$^{(5)}$}
\mbox{J.D. Reid                     \unskip,$^{(10)}$}
\mbox{A. Rimai                      \unskip,$^{(8,\ddag\ddag)}$}
\mbox{A. Rose                       \unskip,$^{(10)}$}
\mbox{F.S. Rotondo                  \unskip,$^{(12,\S\S)}$}
\mbox{J. Sandweiss                  \unskip,$^{(12)}$}
\mbox{R.P. Scharenberg              \unskip,$^{(8)}$}
\mbox{A.J. Slaughter                \unskip,$^{(12)}$}
\mbox{G.A. Smith                    \unskip,$^{(7)}$}
\mbox{M.L. Tincknell                \unskip,$^{(8,\|\|)}$}
\mbox{W.S. Toothacker               \unskip,$^{(7)}$}
\mbox{G. Van Buren                  \unskip,$^{(6)}$}
\mbox{F.K. Wohn                     \unskip,$^{(4)}$}
\mbox{Z. Xu                         \unskip,$^{(12)}$}
\it
\vskip \baselineskip
\centerline{(The E864 Collaboration)}
  $^{(1)}$ Brookhaven National Laboratory, Upton, New York 11973 \break
  $^{(2)}$ University of California at Los Angeles, Los Angeles, California 90095 \break  
  $^{(3)}$ University of California at Riverside, Riverside, California 92521 \break
  $^{(4)}$ Iowa State University, Ames, Iowa 50011 \break 
  $^{(5)}$ University of Massachusetts, Amherst, Massachusetts 01003 \break 
  $^{(6)}$ Massachusetts Institute of Technology, Cambridge, Massachusetts 02139 \break 
  $^{(7)}$ Pennsylvania State University, University Park, Pennsylvania 16802 \break 
  $^{(8)}$ Purdue University, West Lafayette, Indiana 47907 \break 
  $^{(9)}$ United States Military Academy, West Point \break
  $^{(10)}$ Vanderbilt University, Nashville, Tennessee 37235 \break 
  $^{(11)}$ Wayne State University, Detroit, Michigan 48201 \break 
  $^{(12)}$ Yale University, New Haven, Connecticut 06520 \break
\end{center}

\newpage

\date{\today} \maketitle
\begin{abstract}
We present results from Experiment 864 for antiproton production and
antideuteron limits in Au + Pb collisions at 11.5 GeV/c per nucleon.  We
have measured invariant multiplicities for antiprotons for rapidities
$1.4<y<2.4$ at low transverse momentum as a function of collision geometry.  
When compared with the results from Experiment 878 our measurements
suggest a significant
contribution to the measured antiproton yield from the decay of
strange antibaryons.  We have also searched for antideuterons and see
no statistically significant signal.  Thus, we set upper limits on the
production at approximately $3 \times 10^{-7}$ per 10\% highest
multiplicity  Au + Pb interaction.  
\end{abstract}

 \narrowtext

\section{Introduction}

Antimatter production in relativistic heavy ion collisions has been
proposed as an excellent probe of the collision dynamics and possible
phase transition to a quark-gluon plasma \cite{qgprefs}.  The production of
antiprotons at AGS energies (10-15 GeV/c)  is near threshold in
nucleon-nucleon collisions.  Therefore multiple collisions, resonance
excitations, and collective effects may play a major role in significantly
increasing the overall production rates \cite{mean}.  
Strange antibaryons, due to
their larger mass and additional antistrange quark, are even further
suppressed in initial nucleon-nucleon collisions. 
However, strangeness saturation and antimatter enhancement have long been
predictions of quark-gluon plasma states.  Thus, understanding the
yields of non-strange and strange antibaryons is an important tool for
distinguishing between various sources of enhanced production.  

Antibaryons have a large annihilation cross section
(particularly at low relative momentum), reaching levels of hundreds of
millibarns.  Thus, in these baryon rich colliding systems, there can be
significant annihilation losses before they escape the collision region.
The final experimentally measured yields represent the initial
production minus the annihilation losses.  The annihilation process in
free space is well understood and experimentally parametrized;
however, this annihilation may be modified in the dense particle
environment where the initial attraction of baryon and antibaryon may
be disturbed \cite{arc_pap}.  

Antideuteron production at AGS energies is actually below the energy
threshold in single nucleon-nucleon collisions and thus antideuterons are
expected to only be created through the coalescence of separately
produced antiprotons and antineutrons which are close enough together
in coordinate space and phase space at freeze-out.  Because of the large energy
required for their production, antideuteron yields are an excellent
measure of the system's thermal temperature (assuming antimatter is
equilibrated in these collisions).  Also, coalescence yields would yield
information about the spatial distribution of antinucleons (as a form of
two particle correlation) \cite{Nagle_prl}.

In the next section we describe the E864 spectrometer and the 
antiproton data sets. In section three we present the measured
antiproton invariant multiplicities and compare them with 
measurements made by other experiments.  These comparisons lead
us to consider the possibility of enhanced production of strange
antibaryons in these collisions.  Antimatter correlations
in the form of events with two antiprotons and antideuteron
production are discussed in section four.

\section{Experiment 864}

\subsection{The E864 Spectrometer}

\begin{figure}
\vspace{4cm}
\centerline{\hbox{\psfig{figure=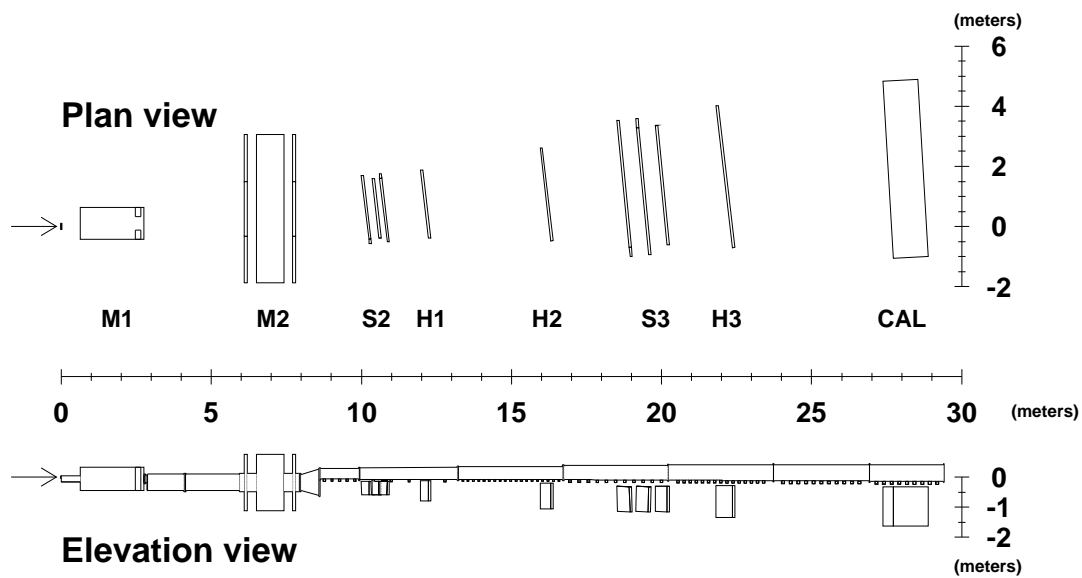,height=7.5cm}}}
\vspace{1cm}
\caption{The E864 detector in plan and elevation views, showing the 
dipole magnets M1 and M2, 
hodoscopes (H1, H2 and H3), straw tube arrays (S2 and S3) and 
hadronic calorimeter (CAL). The vacuum chamber is not shown in the 
plan view.}
\label{fig:e864}
\end{figure}

Experiment 864 was designed to search for novel forms of matter
(particularly strange quark matter, or ``strangelets'') 
produced in heavy ion collisions
at the Brookhaven AGS facility \cite{e864_nimpap}.  
In order to conduct this search,
E864 has a large geometric acceptance 
and operates at a high data rate. A diagram of the spectrometer is
shown in Figure \ref{fig:e864}.
Secondary particles produced from the Au + Pb reaction which are
within the geometric acceptance traverse two dipole magnets (M1 and M2)
and then multiple downstream tracking stations.  Non-interacting beam
particles and beam fragments near zero degrees pass above the downstream
spectrometer in a vacuum chamber, thus reducing interactions which
would otherwise produce background hits in the detectors.  
The experiment does not measure at zero 
degrees (zero transverse momentum), but particles with at least 15 
milliradians of angle pass through an exit window in the vacuum
chamber.  

Charged particles are tracked using three time-of-flight (TOF)
hodoscopes and two straw tube stations.  The hodoscopes yield
three independent charge measurements of dE/dx over the 1 cm
thickness of the scintillator slats and provide three 
space-time points with time resolutions on the order of 120-150 ps.  
The straw stations provide more accurate position information
for track projection back into the magnetic field region.  Particles
are identified by their time-of-flight from the target (yielding
the velocity) and momentum.  The momentum is determined by combining
the charge measurement with the rigidity (R = p/Z) from the track
projection in the bend plane into the field region.  The redundant
measurements allow for excellent background rejection of
ghost tracks and tracks originating from downstream interactions.

A second particle identification measurement can be made using a hadronic calorimeter
located at the end of the spectrometer \cite{calo_nim}.  
The calorimeter consists of
754 individual towers constructed from lead sheets with scintillating fibers
running almost parallel to the incident particle trajectory.  The calorimeter
yields good timing information ($\sigma \approx$ 400~ps for hadrons) 
and excellent hadronic energy resolution of 3.5\% + 34\%/$\sqrt{E}$ (with E in GeV).
For baryons, the calorimeter measures the particle's kinetic energy, which 
when combined with time-of-flight information gives a measure of the
particle mass.  For antibaryons, the energy measurement also includes the 
annihilation energy of the antibaryon and its annihilation partner.

The experiment is able to perform high sensitivity searches by running at
high rate with a special ``late energy'' trigger (LET) \cite{let_nimpap}.  
The time and energy signals from each of 616 fiducial calorimeter 
towers are digitized in flash ADC's and TDC's and used as inputs to a lookup table, which
is programmed to select the particles of interest.
Because there are many slow neutrons and many fast high energy protons, a
simple time cut or an energy cut was determined to be insufficient
for the trigger.  The late energy trigger allows for the rejection
of both of these abundant particles, while effectively
triggering on slow (mid-rapidity) particles which deposit
a large amount of energy.  
An antiproton of the same momentum as
a proton or neutron will deposit an additional annihilation energy.   
If the Au+Pb interaction yields no towers
firing the trigger, then a fast clear is sent out to the digitizers and the
data is not recorded.
The trigger yields an enhancement
factor for antiprotons, antideuterons and strangelets of approximately 50
(under running conditions appropriate for each species).

In order to determine the collision geometry (impact parameter) 
a charged particle multiplicity counter is used.
The E864 multiplicity counter \cite{beam_nim} is an annular piece of 
fast BC420 scintillator placed around the beam pipe 13 cm downstream 
of the target and tipped 
at an angle of $8^o$ to the vertical.  It is 1 cm thick and 
subtends the angular range of $16.6^o$ to $45.0^o$. The annulus is 
separated into four quadrants and each quadrant is viewed by a 
photomultiplier tube.  The total signal measured with this counter is 
proportional to the charged particle multiplicity of the collision.
The integrated signal from the sum of the four quadrants is
used to trigger on the centrality of the events by selecting events 
with a signal larger than a given threshold.

\subsection{Data Sets}

\begin{figure}
\vspace{4cm}
\centerline{\hbox{\psfig{figure=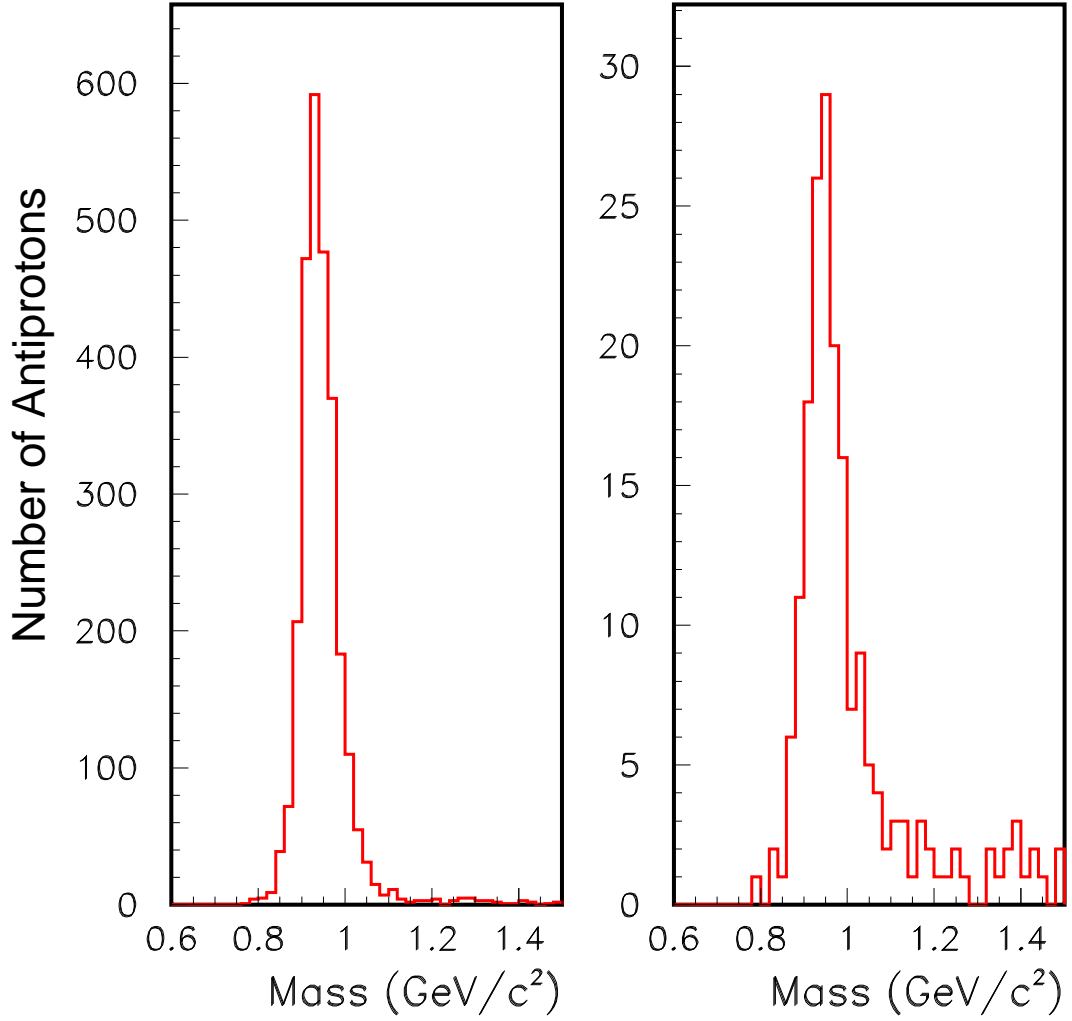,height=13.5cm}}}
\vspace{1cm}
\caption{Antiproton mass distributions for $Z=-1$ particles from the 
1995 (``-0.75T'') data set. The left panel show the mass distribution
for particles that triggered the LET, while the right panel shows
the mass distribution for particles that did not. These mass distributions are
for particles with rapidity $1.8<y<2.0$ and transverse momentum 
$125< p_{T}< 150$ MeV/c. }
\label{fig:mass_plots}
\end{figure}

The data used in this analysis was collected in two separate data taking periods.
During the fall of 1995, the late energy trigger was strobed on
the 10\% most central Au+Pb interactions with the spectrometer
magnets set for optimal acceptance for antideuterons (referred to as the ``-0.75T'' field setting).
The LET curve was set to yield an enhancement factor of
$\sim50$ for antideuterons and negative strangelets.  The data set includes over 
90 million recorded events, which effectively sampled approximately six
billion central interactions.   From this sample, over 50,000 antiprotons were 
identified.  The mass distributions of antiprotons from a single rapidity and
transverse momentum bin are shown in Figure \ref{fig:mass_plots}.  

In the fall 1996 run, the LET was strobed on minimum-bias (93\% of the geometric 
cross section) Au+Pb 
interactions and the LET curve and the spectrometer magnets were set 
for optimal antiproton acceptance (referred to as the ``-0.45T'' field setting). The LET yielded an 
enhancement factor for antiprotons $>50$ under these conditions.  However, in
order to use the trigger effectively, the region of the 
calorimeter dominated by neutrons from the interaction
had to be excluded.
This reduced the geometric acceptance by roughly a factor of two.
The data sample included 45 million recorded minimum bias interactions and 
approximately 50,000 antiprotons.  These data samples represent the largest 
statistics for antiprotons produced in heavy ion collisions at the BNL-AGS.

In both data sets, the beam momentum was measured using the E864 spectrometer magnets and a 
downstream beam counter located in the beam dump.  The beam momentum of 11.5 GeV/c per nucleon
was consistent with the beam momentum reported at extraction from the accelerator once energy
losses due to material in the E864 beam line were properly accounted for.  The Au beam was 
incident on a 30\% Pb target for the 1995 data set, while a 10\% Pb target was used in 1996.

\section{Antiproton Invariant Multiplicities}

\subsection{E864 Measurements}

In E864 we explicitly measure the yield of antiprotons per
Au+Pb interaction as a function of centrality, and thus
we directly calculate the invariant multiplicities.   
The invariant multiplicity for antiprotons is
determined as follows:
\begin{equation}
\rm {{{1} \over {2 \pi p_{t}}} {{d^{2}N} \over {dydp_{T}}}} = {{1}
\over {2\pi \overline{p_{t}} \Delta y \Delta p_{T} }} {{N_{counts}}
\over {N_{sampled}}} {{1} \over {\epsilon_{detect} \times
\epsilon_{accept} \times \epsilon_{trigger}}} 
\end{equation}
The total number of antiprotons $N_{counts}$ is
determined in each separate bin in phase space and divided by the
total number of sampled Au+Pb interactions $N_{sampled}$.
The counted antiprotons include only those antiprotons which fired the LET.  
Since the detector does not measure all the
antiprotons produced in a given region of phase space, the
invariant multiplicity must be corrected for the missed particles.
These missed antiprotons are the result of the experiment's finite
acceptance and various tracking and triggering efficiencies.  The
acceptance $\epsilon_{accept}$ and detection efficiency
$\epsilon_{detect}$ are calculated using a GEANT \cite{GEANT} simulation of the
experiment in conjunction with real data.  This simulation also included 
losses due to antiproton annihilation in the target as part of the acceptance correction. 
The production of antiprotons due to reinteraction of particles from the 
primary interaction with target nuclei was also considered and found to 
be negligible.

The LET efficiency $\epsilon_{trigger}$ is determined in
each kinematic bin.  This efficiency is determined
in one of two ways: for antiprotons measured in the 1995 (``-0.75T'') 10\% central data
where the efficiency was somewhat low, a sample of antiprotons that did 
not fire the trigger was used to determine the efficiency.  In the
1996 (``-0.45T'') data, where the LET curve was set for higher efficiency
($\sim $ 75\%), the efficiency was determined from a Monte Carlo
of the calorimeter response.

\begin{figure}
\vspace{4cm}
\centerline{\hbox{\psfig{figure=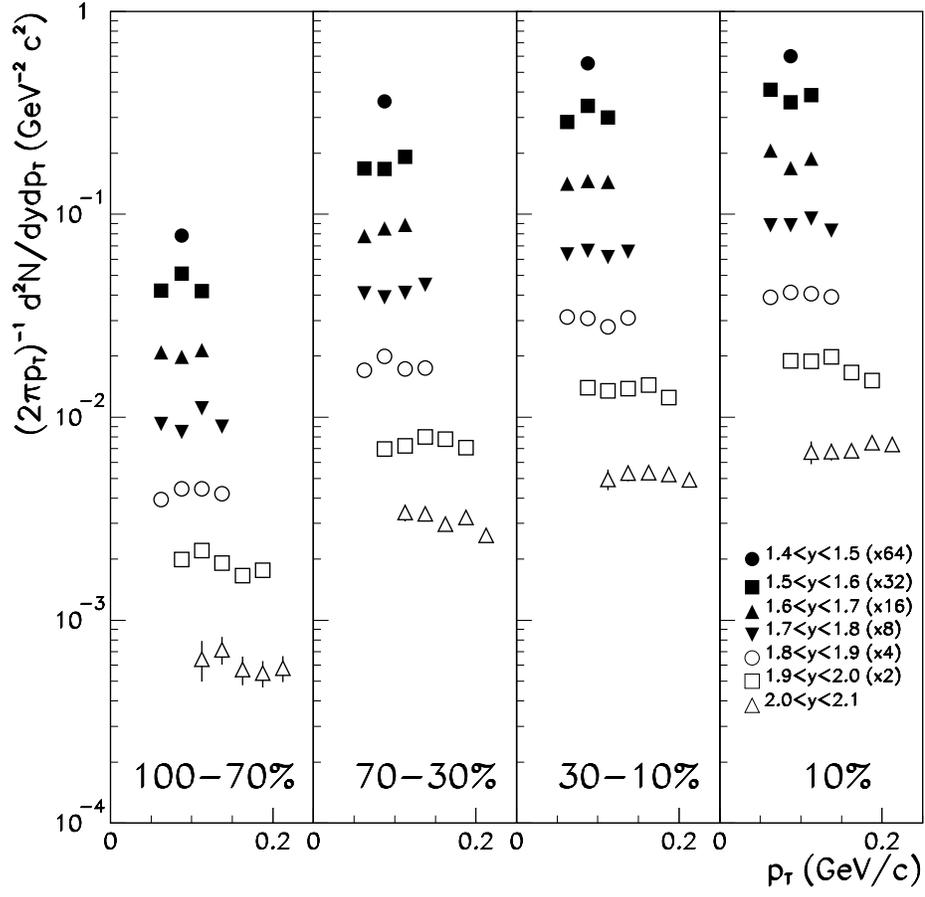,height=13.5cm}}}
\vspace{1cm}
\caption{Invariant multiplicities for antiprotons as measured in the 
four different centrality bins from the 1996 data.  Note that the
data are approximately flat in each rapidity interval over the
measured range in transverse momentum.}
\label{fig:pt_plots}
\end{figure}

The data from E864 is measured in a range
of $50~<~p_{T}~<~275~$MeV/c (where the limits are a function
of rapidity).  The invariant multiplicities
measured in E864 are approximately flat over the $p_{T}$ range
measured, as shown for the 1996 data in Figure \ref{fig:pt_plots}.  
Over such a small range in transverse
momentum, the invariant multiplicities are not expected to change
significantly.  If, for example, the spectra follow a Boltzmann
distribution,
\begin{equation}
{{1} \over {2\pi p_{T}}} {{d^{2}N} \over {dydp_{T}}} \propto m_{T}
e^{-{{m_{T}} \over T_{B}}}
\end{equation}
(where $m_{T} = \sqrt{p_{T}^{2} + m^{2}}$), then with a temperature
parameter of 200~MeV the invariant multiplicity at $p_{T}=0$ is only
6\% higher than at $p_{T}~=~150$~MeV/c. For comparison with other 
experiments, in each rapidity bin all
the invariant multiplicities as a function of $p_{T}$ 
are fit to a constant level.  This level
is assigned as the invariant multiplicity at $p_{T} \approx 0$, and 
an additional 6\% systematic error is assigned due to the $p_{T}=0$ 
extrapolation. It should be noted that strong radial flow could affect the
$p_{T}=0$ extrapolation as well. We feel that this effect
should be within the estimated systematic uncertainty since the E864 data
are quite flat as a function of $p_{T}$ down to 50 MeV/c at midrapidity. 

The antiproton invariant multiplicities for 10\% most central Au + Pb
collisions from the 1995 data are given in Table \ref{tab:pbar_invar95} \cite{nagle_thesis}.  
It should be noted that the statistical error in this data set 
is dominated by the contribution from the trigger efficiency (due to
counting antiprotons which did not fire the trigger). The systematic
error in the 1995 data (exclusive of the 6\% due to the $p_{T}=0$ extrapolation)
is estimated to be 15\%, and is dominated by the
uncertainty in the correction for the LET trigger efficiency. Systematic 
uncertainties also arise from our knowledge of the experimental acceptance
(including the effect of the collimator in the first spectrometer magnet),
track quality cuts, and the loss of tracks due to overlapping hits in the 
hodoscopes.

\begin{figure}
\vspace{4cm}
\centerline{\hbox{\psfig{figure=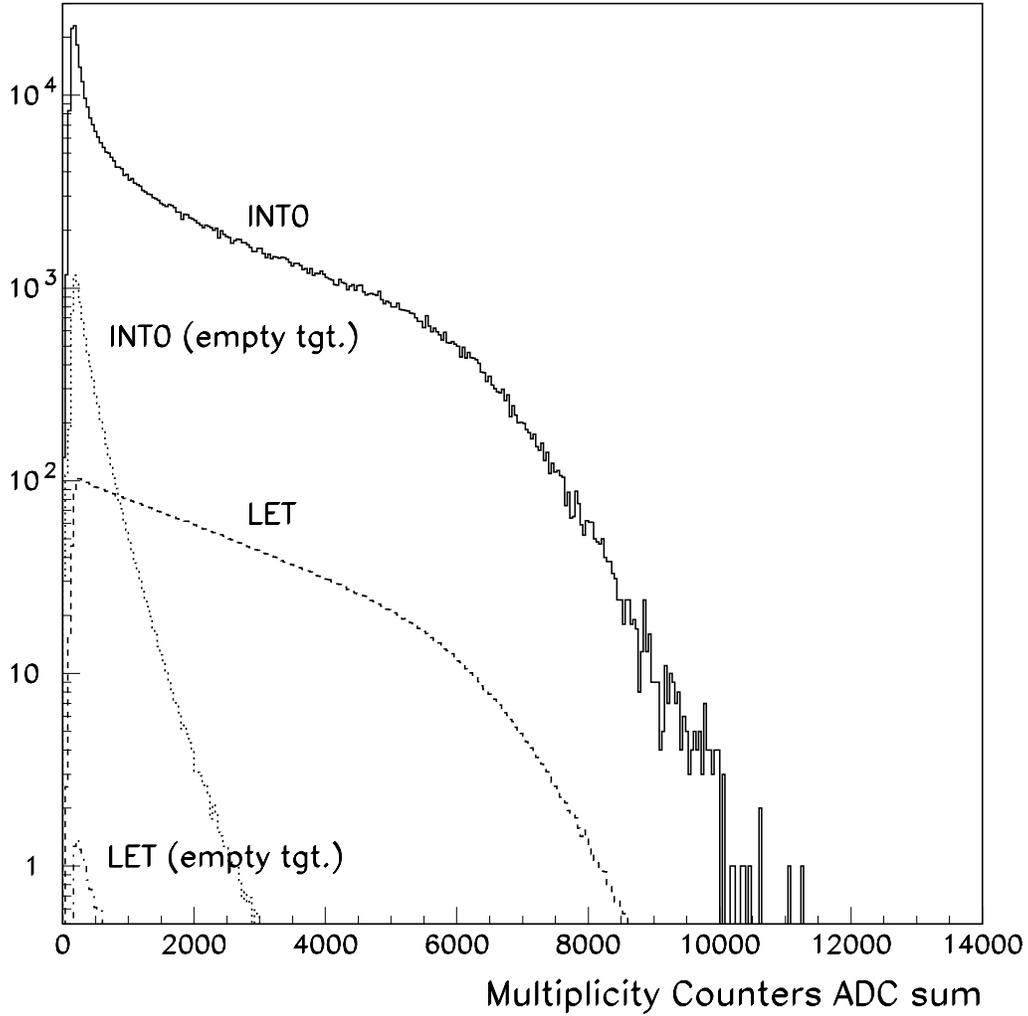,height=13.5cm}}}
\vspace{1cm}
\caption{Multiplicity counter ADC distributions for INT0 (minimum bias) and LET triggers
for 10\% Pb and empty targets; the distributions are scaled to the same number
of incident Au ions.  For LET triggers the empty target
contribution is less that 1\% of the distribution for the lowest 
multiplicities.}
\label{fig:beam1}
\end{figure}

\begin{figure}
\vspace{4cm}
\centerline{\hbox{\psfig{figure=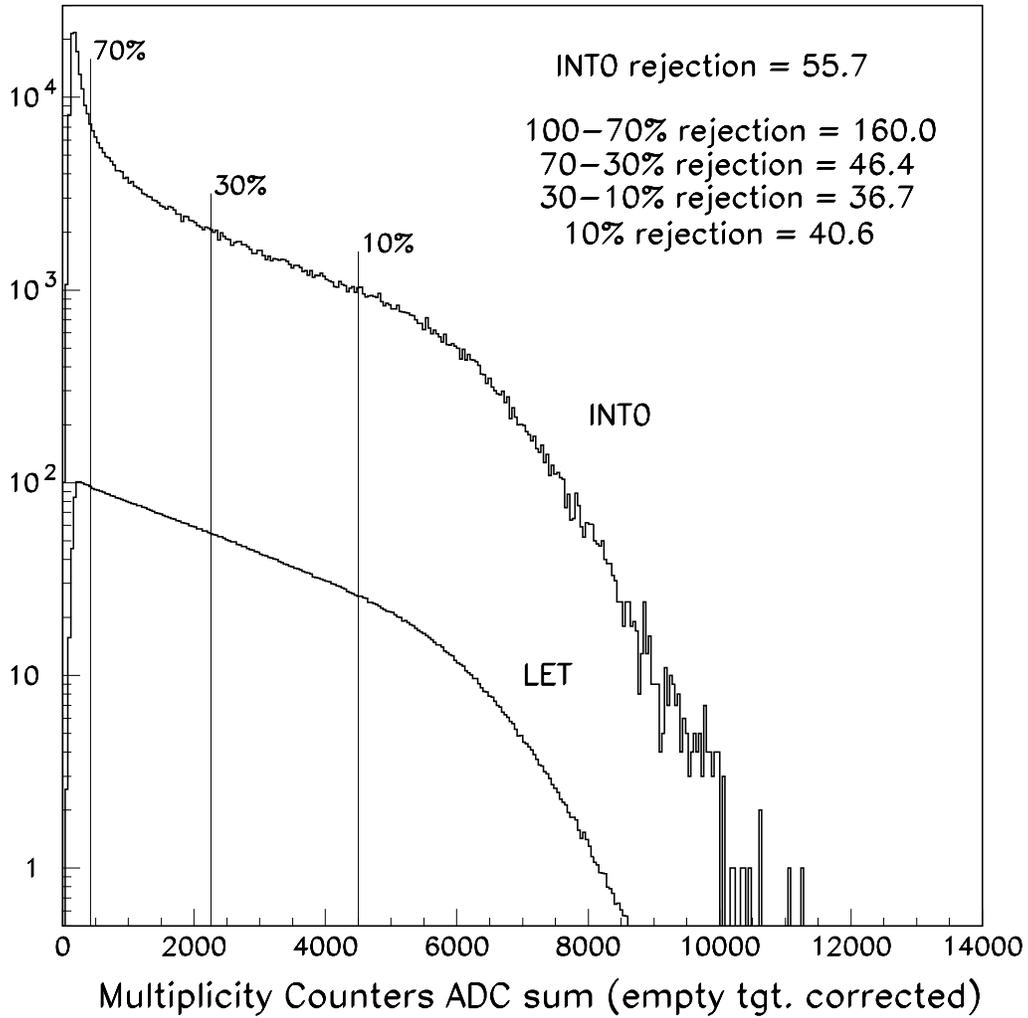,height=13.5cm}}}
\vspace{1cm}
\caption{Multiplicity counter ADC distributions (corrected for empty 
target contributions) showing the centrality
cuts used in the analysis.  The effective LET rejection factor 
for each multiplicity region is also shown.}
\label{fig:beam2}
\end{figure}

For the 1996 data, the late-energy trigger was strobed on a minimum-bias
sample of events selected by the multiplicity counter. The resulting 
multiplicity counter ADC distribution is shown in Figure \ref{fig:beam1}.
When selecting minimum bias events, it is important to consider the
effect of interactions of the beam that do not occur in the target. 
Using special empty-target runs we have found that non-target interactions
contribute less than 10\% of the multiplicity distribution at low 
multiplicity, while the late-energy trigger further reduces this 
contamination to below 1\% (see Figure \ref{fig:beam1}).  For the 
1996 data, the antiproton invariant multiplicities are determined
for different regions of the minimum bias multiplicity: 100-70\% of the full
distribution, 70-30\%,
30-10\% and 10\%.  These regions are shown in Figure \ref{fig:beam2}.
It is important to note that the LET rejection is a strong function
of the multiplicity, and this must be properly accounted for when 
calculating the normalization in each centrality bin.  

The antiproton invariant multiplicities for the four multiplicity
regions used in the 1996 data set are listed in Table \ref{tab:pbar_invar96}. 
In addition, the full minimum bias invariant multiplicity from the 1996 data
set is also listed in Table \ref{tab:pbar_invar96}.      
The systematic error in these data points are estimated to be 10\% (again,
exclusive of the 6\% previously described due to the $p_{T}=0$ extrapolation). 
As in the 1995 data, the systematic uncertainty in the 1996 data is also dominated
by the uncertainty in the determination of the LET trigger efficiency.
However, the size of the correction is smaller for the 1996 data due to
the overall higher efficiency of the trigger setting.  

\begin{figure}
\vspace{4cm}
\centerline{\hbox{\psfig{figure=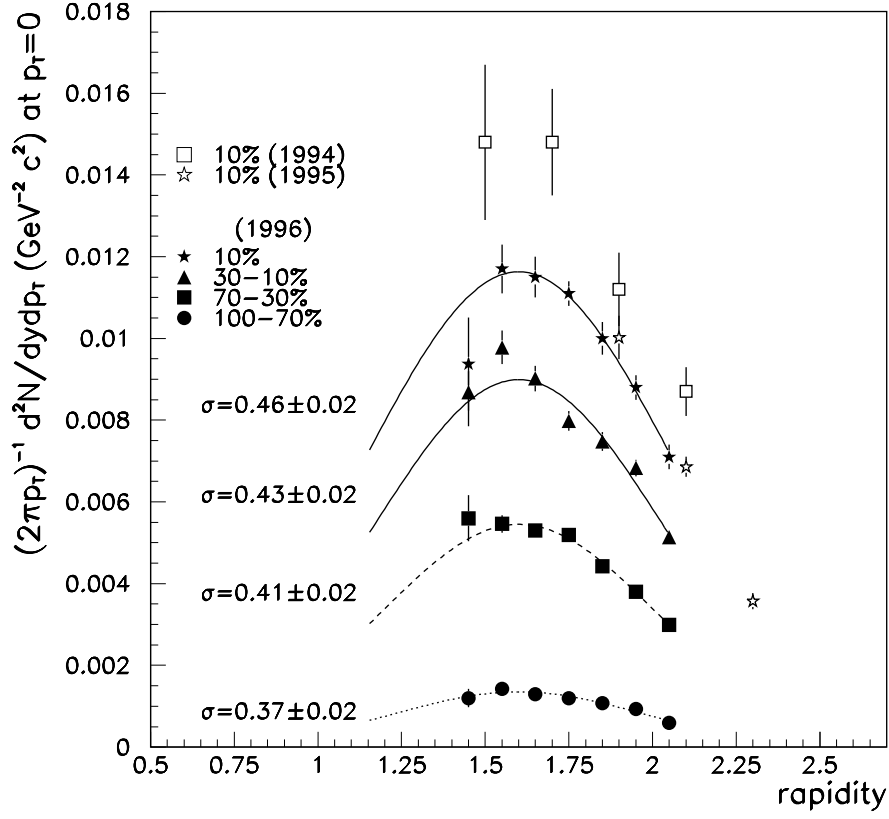,height=13.5cm}}}
\vspace{1cm}
\caption{Antiproton invariant multiplicities as a function of centrality showning
the 1994, 1995 and 1996 data sets from E864. The error bars shown are 
statistical only. Systematic errors 
are estimated to be 20\% on the 1994 data, 15\% for the 1995 data, 
and 10\% for the 1996 data, not including a 6
to $p_{T}=0$. 
The fits are constrained to have a mean value at midrapidity ($y=1.6$).} 
\label{fig:fig_all_years}
\end{figure}

Figure \ref{fig:fig_all_years} shows the 1995 (``-0.75T'') 
and 1996 (``-0.45T'') antiproton invariant multiplicities at $p_{T}=0$
as a function of rapidity.  The Gaussians shown are fits to the
combined 1995 and 1996 data, and are constrained to have a mean value at 
midrapidity ($y=1.6$).  There is excellent agreement between the 
1995 and 1996 data in the rapidity range where the two data sets overlap.
The rapidity widths measured in the data for the four multiplicty
widths are $0.37\pm0.02$ (100-70\%), $0.41\pm0.02$ (70-30\%),
$0.43\pm0.02$ (30-10\%) and $0.46\pm0.02$ (10\%), indicating a broadening
of the rapidity spectrum at higher centrality.

Also shown in Figure \ref{fig:fig_all_years} (as open squares) are previously
reported antiproton results from data taken in 1994 (``-0.45T'') \cite{jlprl}.  
It should be noted that the 1994 data is about
20\% higher at midrapidity than indicated by the corresponding 
1995 and 1996 data.  This is within the statistical and systematic
error previously quoted for the 1994 data.  It is important 
to note that the 1994 data was taken with an incomplete detector:
two layers of S3 were missing along with a ``plug'' designed
to reduce the occupancy in the downstream detectors due to 
interactions of beam-rapidity fragments with the vacuum chamber.
The presence of the plug dramatically reduced the detector
occupancy in the 1995 and 1996 data (and hence the size of the
correction required for losses due to multiple hodoscope hits),
and the presence of the additional S3 layers provided additional 
background rejection.

\subsection{Comparisons with Other Experiments}

Experiment 878 has measured antiproton yields as a function of
collision geometry in reactions of Au+Au ions at
10.8~A~GeV/c \cite{e878_prl,mike_prc}.   There are two
differences between the reaction system between E878 and E864:  (1) the
target in E864 is Pb and (2) the beam momentum in E864 is higher at
11.5~A~GeV/c.  The target difference is quite small and is neglected
in this comparison.  However, the production of antiprotons is near
threshold for nucleon-nucleon collisions at these energies and so the
beam momentum difference must be accounted for.  We assume that the
ratio of antiproton yields in Au+Pb reactions at the two energies is
proportional to the ratio of antiproton yields in p+p reactions at the
two energies.   Unfortunately, there is no usable data on antiproton
production in p+p reactions covering this  particular energy range.
Therefore a parametrization of the production cross sections (derived
from p+p data at higher energies and p+Be data from E802 at
14.6~GeV/c \cite{arc_pap}) is employed.  
Using this parameterization, one expects the ratio of the antiproton production
cross sections at the two energies to be 1.5. 
The E878 invariant multiplicities are scaled up by this value.  
By considering fits to higher energy p+p data that do not include the E802 p+Be data
at 14.6 GeV/c \cite{costales}, 
we estimate that 
this energy scaling contributes an additional 15\%
systematic error on the overall normalization of the scaled E878 points. 

\begin{figure}
\vspace{4cm}
\centerline{\hbox{\psfig{figure=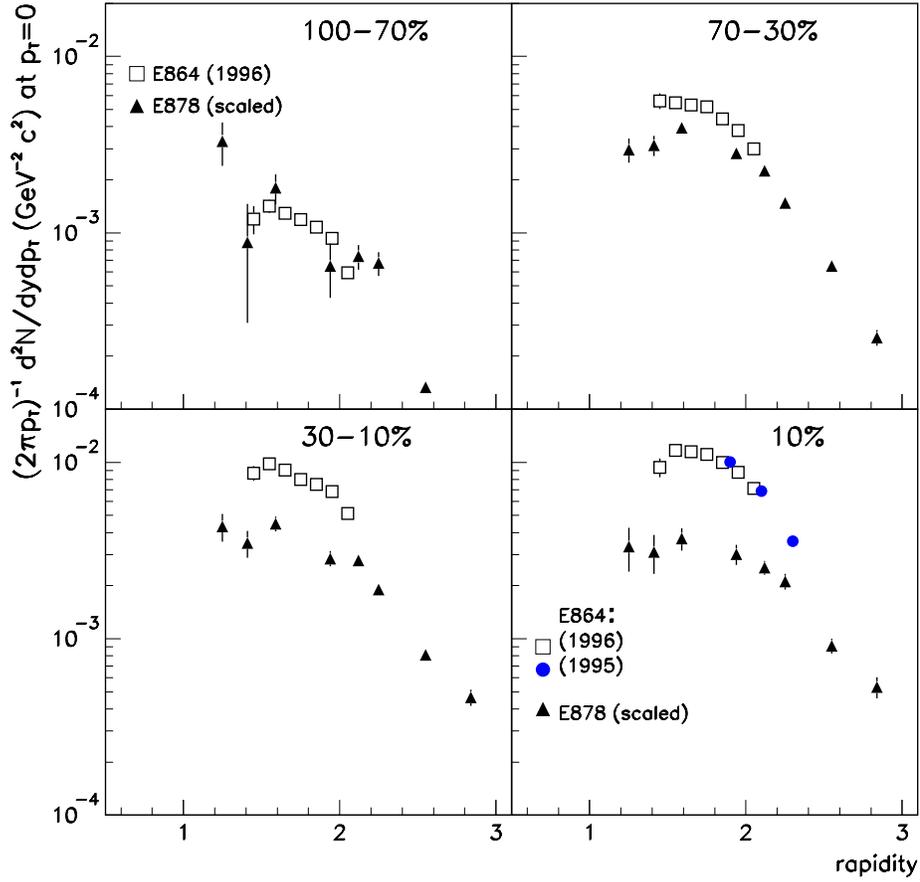,height=13.5cm}}}
\vspace{1cm}
\caption{Antiproton invariant multiplicities at $p_{T}=0$ compared to E878
as a function of centrality.  Note that while the two experiments agree
for peripheral collisions, they disagree markedly for more central
collisions.  The E878 data have been scaled up as described in the
text to account for the lower beam momentum.} 
\label{fig:fig_pbar_e878_comp}
\end{figure}

Experiment 878 measures invariant multiplicities nominally at
$p_{T}=0$ (which is really at transverse momenta less than
$\sim$ 30-50~MeV/c). Using the procedure previously 
outlined, we extrapolate the E864 measurements to $p_{T}=0$
and compare the E864 and E878 measurements in 
Figure \ref{fig:fig_pbar_e878_comp}.  While the two experiments agree well for low 
multiplicity collisions a substantial disagreement develops
for more central collisions.  For 10\% central collisions,
the E864 measurements at midrapidity are a factor of $\sim3.2$ larger
than the corresponding E878 data.



It should be noted that both experiments do not use precisely the
same definition of centrality: E864 measures the multiplicity 
of charged particles produced in the collisions, while E878 
measures the $\gamma$ multiplicity (mostly from $\pi^{0}$ decay).  In order to properly compare
the two experiments the multiplicity ranges for both experiments must be converted 
to a (somewhat model-dependent) parameter.  To do this, we have chosen
to show the integrated antiproton yield at $p_{T}=0$ versus the number of ``first''
nucleon-nucleon collions in each centrality range.
In order to estimate the number of first collisions in each 
multiplicity range for the E864 data, a GEANT \cite{GEANT} simulation was used
in conjunction with RQMD \cite{RQMD} Au+Pb events to generate a trigger probability
vs. impact parameter distribution for each multiplcicity region (see Figure \ref{fig:bdists}).
These distributions were then folded with distributions of the number
of first collisions vs. impact parameter from a simple Glauber
model calculation.  A similar procedure was applied to the E878
data, using the results of a simulation of the E878 multiplicity array \cite{e878_mult_nim}.  The results
of this exercise (shown in Figure \ref{fig:first}) demonstrates that 
the E864 and E878 centrality ranges are quite similar.   

\begin{figure}
\vspace{4cm}
\centerline{\hbox{\psfig{figure=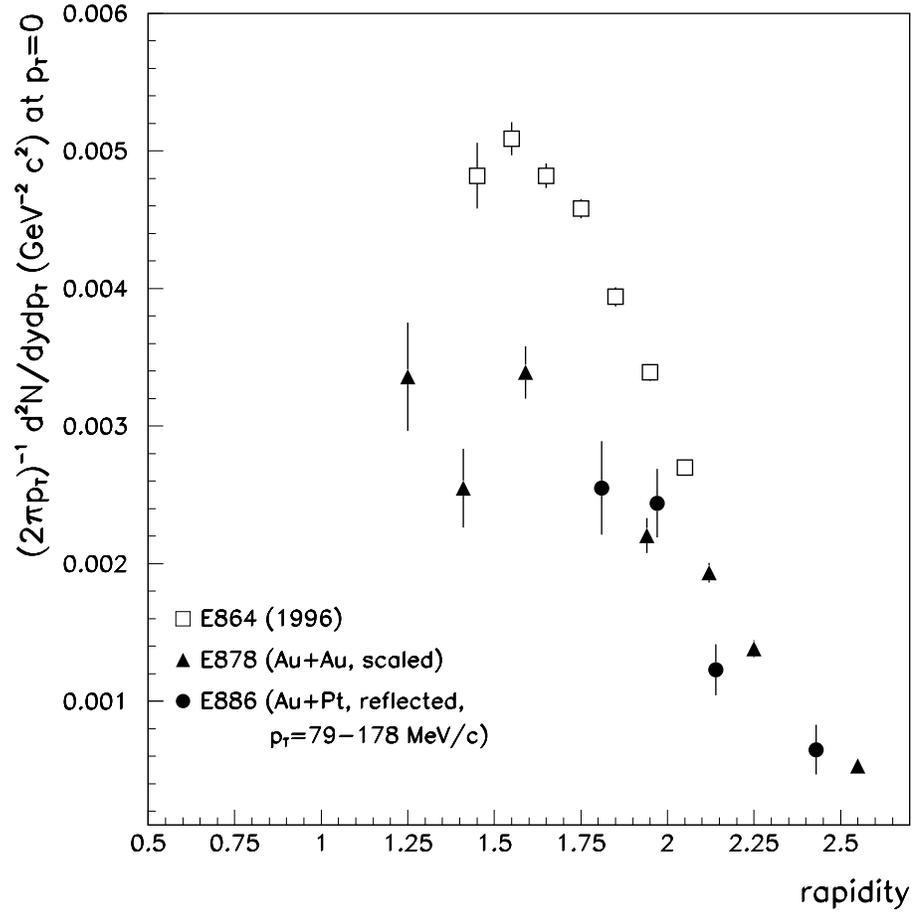,height=13.5cm}}}
\vspace{1cm}
\caption{Antiproton invariant multiplicities from the 1996 (-0.45T) minimum bias data.
Also shown is minimum bias data from E878 (scaled up to 11.5 GeV/c)
and data from E886 (shown reflected about midrapidity, $y=1.6$).} 
\label{fig:fig_pbar_mbias}
\end{figure}

\begin{figure}
\vspace{4cm}
\centerline{\hbox{\psfig{figure=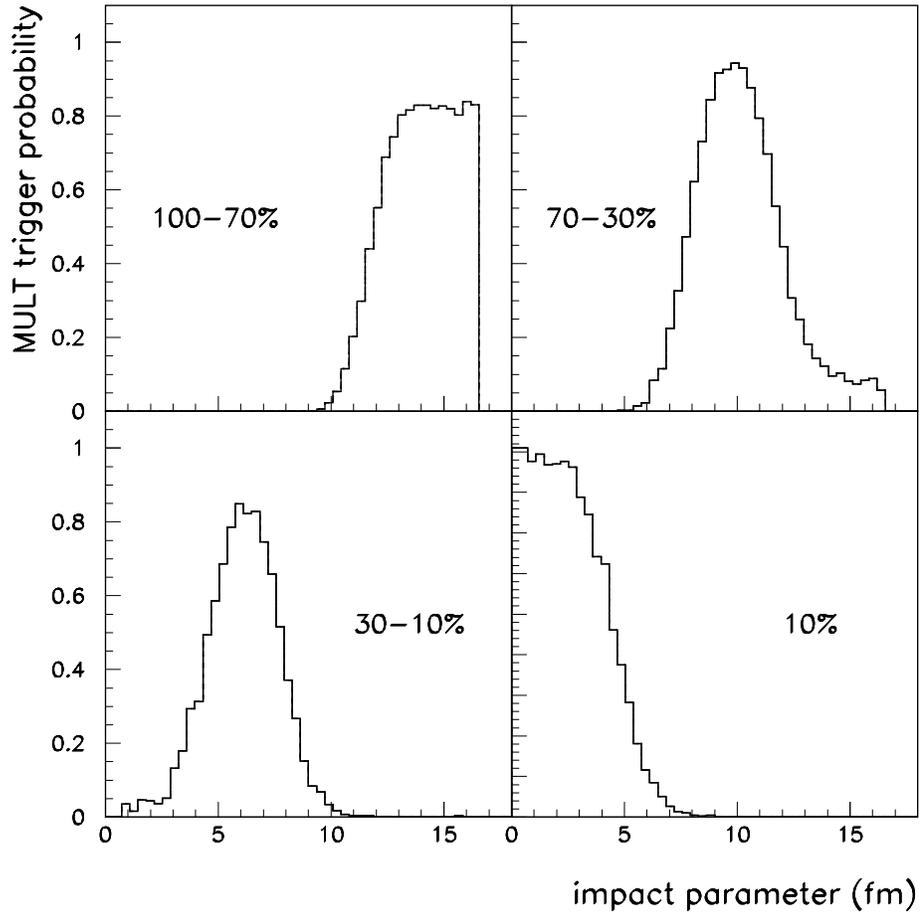,height=13.5cm}}}
\vspace{1cm}
\caption{Trigger probability vs. impact parameter for the E864 multiplicity counter
in the four centrality regions used in the analysis of the 1996 data.  These
distributions were generated using RQMD events in a GEANT simulation of the
E864 multiplicity counter.} 
\label{fig:bdists}
\end{figure}

\begin{figure}
\vspace{4cm}
\centerline{\hbox{\psfig{figure=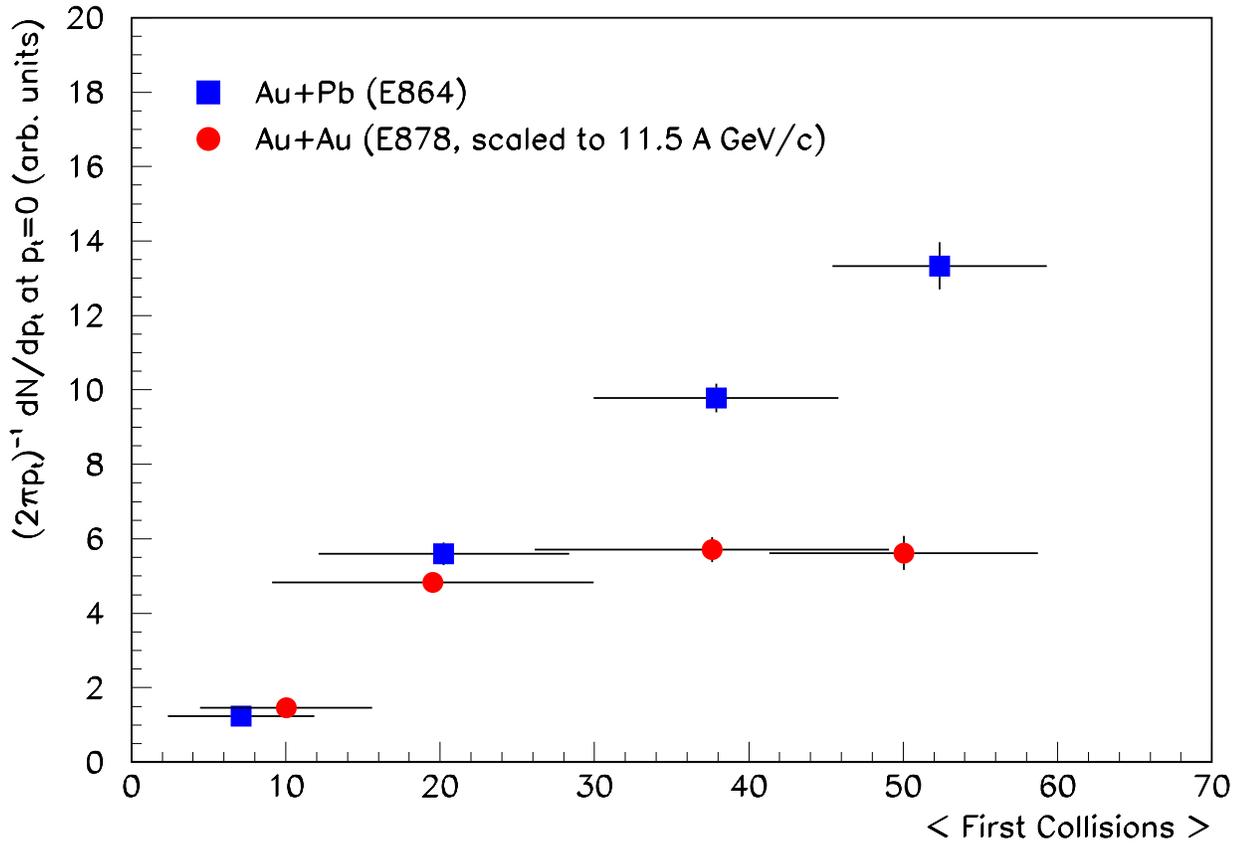,height=13.5cm}}}
\vspace{1cm}
\caption{Integrated antiproton yield at $p_{T}=0$ vs. the mean number of first collisions.  The number
of first collisions are generated using simulations of the detector response and a Glauber model;
the horizontal bar for each multiplicity range indicates the RMS of the first collisions
distribution.} 
\label{fig:first}
\end{figure}

In Figure \ref{fig:fig_pbar_mbias} we also compare measurements
of the minimum-bias cross section for Au+Pb collisons at 11.5 A GeV/c
with E878 and E886 \cite{e886}.
It should be noted that experiment E886 only measured antiprotons from 
minimum bias collisions and thus there is no comparison as a function of centrality.
As expected by the comparison of the E864 data with E878 for
the four different centrality regions, the minimum bias 
invariant multiplicities measured in E864 are substantially larger than those measured
by E878 and E886.

\subsection{Strange Antibaryon Feed-Down}

There is a scenario which can reconcile the E864 and E878 results.   
Some of the antiprotons measured by the various experiments may be the
daughter product of weak decays of strange antibaryons
($\overline{\Lambda}$, $\overline{\Sigma^{+}}$,
$\overline{\Sigma^{0}}$, $\overline{\Xi^{0}}$, $\overline{\Xi^{-}}$, $\overline{\Omega}$).  
This process is referred to as ``feeding
down'' from the strange antibaryons into the antiprotons.   Due to the
significantly different designs of the two experiments, they have
different acceptances from these decay product antiprotons.  There are
a number of antihyperon ($\overline{Y}$) feed-down channels into the antiproton: 
\begin{equation}
\overline{\Lambda} \rightarrow ~~\overline{p} + \pi
^{+}~~~(\rm{65\%~B.F.}) 
\end{equation}
\begin{equation}
\overline{\Sigma^{0}} \rightarrow ~~\overline{\Lambda} + \gamma
\rightarrow ~~\overline{p} + \pi ^{+} + \gamma~~~(\rm{100\% \times
65\%~B.F.}) 
\end{equation}
\begin{equation}
\overline{\Sigma^{+}} \rightarrow ~~\overline{p} + \pi
^{0}~~~(\rm{52\%~B.F.}) 
\end{equation}
\begin{equation}
\overline{\Xi^{0}} \rightarrow ~~\overline{\Lambda} + \pi
^{0} \rightarrow ~~\overline{p} + \pi^{+} + \pi^{0}~~~(\rm{99\%} \times \rm{65\%~B.F.}) 
\end{equation}
\begin{equation}
\overline{\Xi^{-}} \rightarrow ~~\overline{\Lambda} + \pi
^{+} \rightarrow ~~\overline{p} + \pi^{+} + \pi^{+}~~~(\rm{99\%} \times \rm{65\%~B.F.}) 
\end{equation}
and multiple decay modes for the $\overline{\Omega}$.
The decay of the $\overline{\Sigma^{0}}$ will produce additional 
$\overline{\Lambda}$'s which will be 
indistinguishable from those created in the primary collision.  
The decay of the $\overline{\Lambda}$ and the $\overline{\Sigma^{+}}$ will 
produce $\overline{p}$'s whose production vertices do not coincide with the 
location of the primary interaction between
the two nuclei.  Therefore, the degree to which $\overline{p}$'s 
from these decays contribute to a measurement of
$\overline{p}$ production will vary among experiments.  

Due to its large acceptance, the E864 spectrometer will detect 
$\overline{p}$'s from $\overline{Y}$ decay.
E864 does not have sufficient vertical resolution
to distinguish $\overline{p}$'s from $\overline{Y}$ decay based on 
the vertical projection of a particle to the target, 
and the analysis cuts do not preferentially reject 
antiprotons from $\overline{Y}$ decay.
Therefore, the $\overline{p}$'s detected in E864 are a combination 
of primary $\overline{p}$'s and $\overline{p}$'s from $\overline{Y}$ decay, in a ratio that 
reflects their production ratio. 
 
The E878 collaboration have also evaluated the acceptance of their spectrometer for 
feed-down from $\overline{Y}$ decay. 
At midrapidity the acceptance for $\overline{p}$'s 
from $\overline{\Lambda}$ and $\overline{\Sigma^{0}}$ decay is  
14\% of the spectrometer acceptance for primordial $\overline{p}$'s,
and 10\% of the $\overline{p}$ acceptance for $\overline{\Sigma^{+}}$ decays \cite{mike_prc};
the acceptance grows at higher rapidity.  In what follows, we assume a uncertainty 
of $\pm$1\% in the E878 acceptances for feed-down. 

Since both E878 and E864 measure a different combination
of primordial $\overline{p}$ production and feed-down from
$\overline{Y}$ decay, we can in principle separate the
two components if we make two explicit assumptions:
both E864 and E878 understand their systematic errors, and the
entire difference between the two experiments can be 
attributed to antihyperon feed-down.  
It is important to note that in energy scaling the E878 results
we have implicitly assumed that the $\overline{Y}$'s scale
with energy in the same way as the $\overline{p}$'s.    

\begin{figure}
\vspace{4cm}
\centerline{\hbox{\psfig{figure=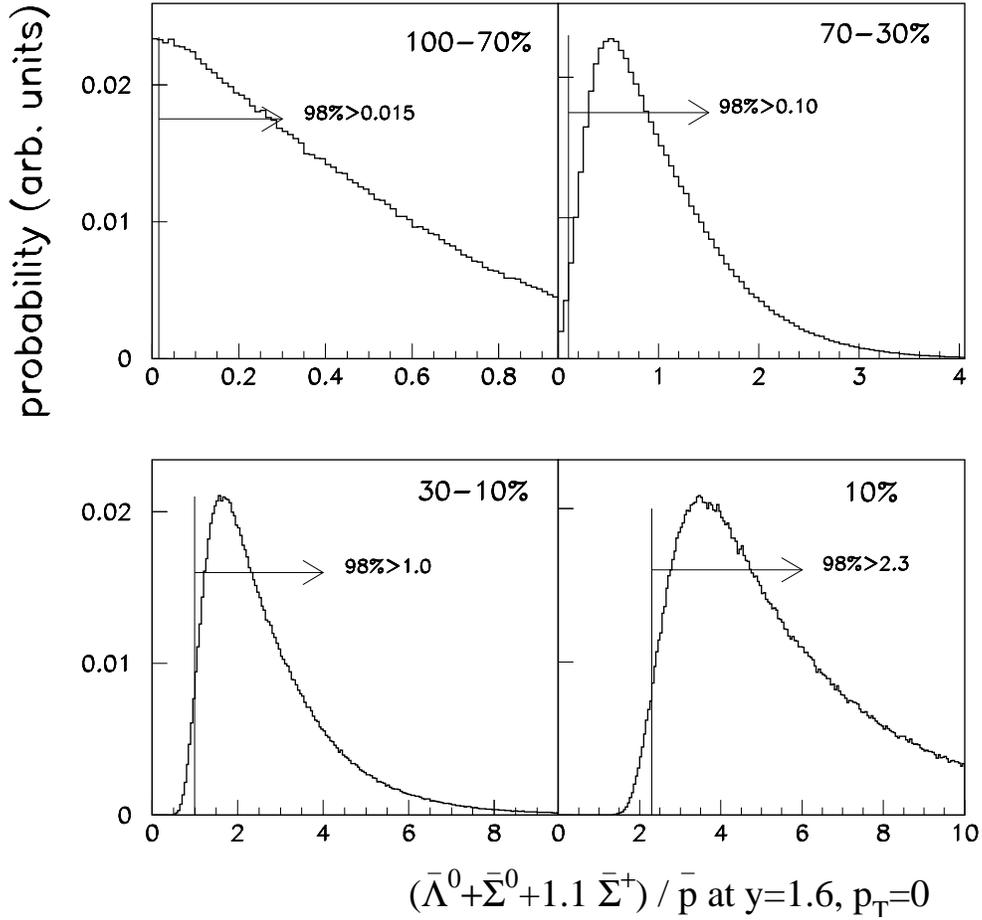,height=13.5cm}}}
\vspace{1cm}
\caption{Probability distributions for the antihyperon/antiproton ratio
as a function of centrality.  The distributions are generated from the ratio
of the E864 antiproton measurements at midrapidity vs. the E878 measurements
using the statistical and systematic errors. See the text for details.  The 
98\% confidence levels are marked in for each distribution.} 
\label{fig:clevel1}
\end{figure}

\begin{figure}
\vspace{4cm}
\centerline{\hbox{\psfig{figure=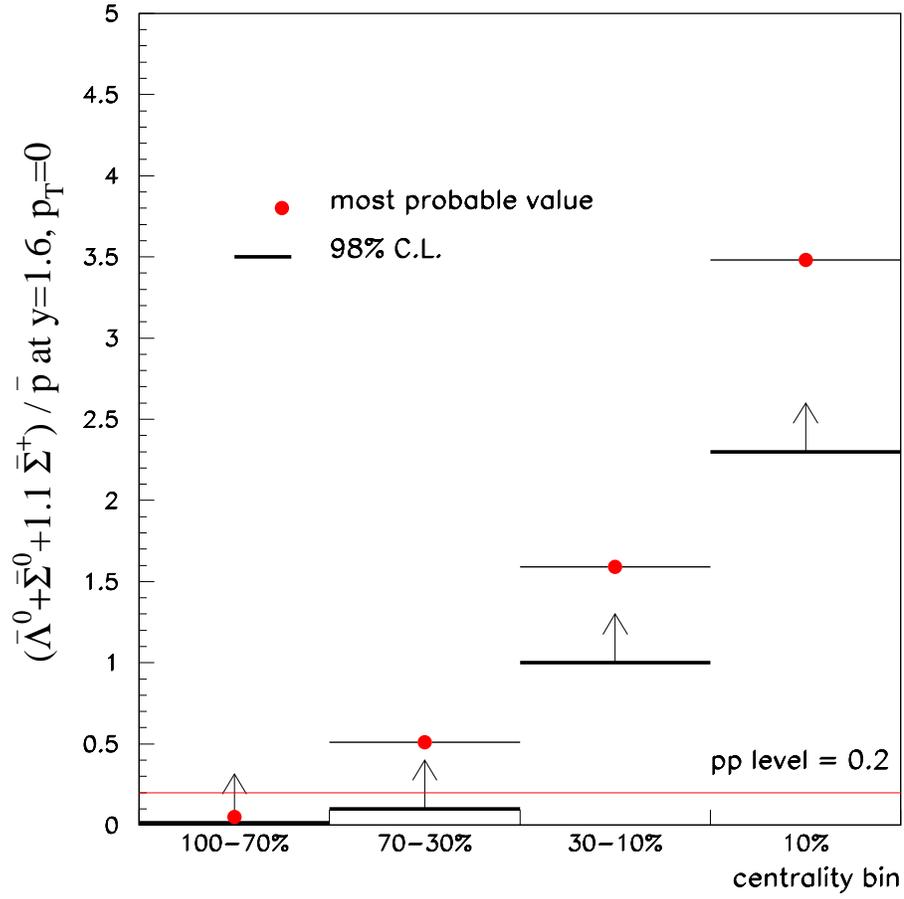,height=13.5cm}}}
\vspace{1cm}
\caption{98\% confidence level values for the antihyperon/antiproton 
as a function of centrality.  The nominal pp level at 12 GeV/c is marked.
Note the dramatic rise in the ratio as a function of collision
centrality.} 
\label{fig:clevel}
\end{figure}

Given an understanding of the errors involved, we can perform
a statistical analysis of the $\overline{Y}/\overline{p}$  
ratio as a function of the E864 and E878 measurements 
(see Figures \ref{fig:clevel1} and \ref{fig:clevel}).
This analysis results in the following limits on the
ratio of $\overline{Y}/\overline{p}$ :
\begin{equation}
\left(\:\frac{\overline{Y}}{\overline{p}}\:\right) 
_{\stackrel{\scriptstyle y=1.6 }{p_{T}=0}} \approx
\left(\:\frac{\overline{\Lambda}+\overline{\Sigma^{0}}
+1.1\overline{\Sigma^{+}}}{\overline{p}}\:\right) >
\textrm{(98\% C.L.)} 
\left\{\: 
\begin{array}{l}
\textrm{0.02 (100-70\%)} \\
\textrm{0.10 (70-30\%)} \\
\textrm{1.0 (30-10\%)} \\
\textrm{2.3 (10\%)} \\
\textrm{0.2 (minimum bias)}
\end{array}
\:\right.
\end{equation}
while the most probable value of this ratio is $\sim3.5$ for 10\% central collisions.
The factor of 1.1 multiplying the $\overline{\Sigma^{+}}$
arises due to the different branching ratio and 
acceptance for the $\overline{\Sigma^{+}}$ compared 
to the $\overline{\Lambda}$.  
The probability 
distributions in Figure \ref{fig:clevel1} were generated
using the measured E864 and E878 invariant multiplicities
for each centraility bin.  The statistical errors on these
measurements were treated as Gaussian, while systematic 
errors on the measurements, energy scaling, $p_{T}=0$ extrapolation, and the 
E878 acceptance for feed-down were treated as definining a limit around the measured
values.

E878 has not explicitly calculated their experimental acceptance for 
the doubly strange $\overline{\Xi}$ and the $\overline{\Omega}$, and thus
they are not explicitly included in the above formula.  These heavier
strange antibaryons are generally thought to be further suppressed and thus
a small contribution.  However, in light of the unexpectedly large ``feed-down''
contributions from strange antibaryons, one should be careful not to
neglect their contribution to this ratio.

This comparison indicates a $\overline{Y}/\overline{p}$ ratio in
Au+Pb collisions that is significantly greater than one at midrapidity 
and $p_{T}=0$.  It should be noted that if the
$\overline{Y}$'s and the $\overline{p}$ are produced with different 
distributions in $y$ and $p_{T}$, then the ratio of integrated yields 
of these particles will differ from the ratio at central rapidity and $p_{T}=0$.
Preliminary results from Si+Au collisions based on 
direct measurements of $\overline{p}$ and $\overline{\Lambda}$ production 
by the E859 collaboration
also indicate a ratio of integrated yields greater than one \cite{Yeudong_Wu}.
For comparison, the $\overline{\Lambda}/\overline{p}$ ratio in pp collisions at 
similar energies is $\sim 0.2$ \cite{pp_ref}. 

An enhancement of antihyperons arises naturally in models that 
include a QGP, and therefore enhanced antimatter and strangeness
production \cite{str_enh_refs,qgprefs}.  Thermal models that use a temperature and baryon chemical 
potential derived from measured particle spectra also indicate
that the $\overline{Y}/\overline{p}$ ratio could be larger than 
one \cite{hgas_refs}.  However, extremely large values of the 
 $\overline{Y}/\overline{p}$ are difficult to achieve in a thermal 
model unless the freezout temperature and/or $K^{+}/K^{-}$ are 
pushed beyond experimentally observed values.  Transport models
such as RQMD \cite{RQMD} predict the $\overline{Y}/\overline{p}$
ratio to be less than one.  Including in a cascade model conversion 
reactions such as
\begin{equation}
\overline{p} + K^{+} \rightarrow \pi + \overline{\Lambda}
\end{equation}
and a lower annihilation cross section for the $\overline{\Lambda}$
relative to the $\overline{p}$ enhances the $\overline{Y}/\overline{p}$
ratio substantially \cite{gerd}.  However, such a model does 
not reproduce the trend with centrality seen in the E864 data.

\section{Antimatter Correlations}

\subsection{Double Antiproton Events}

In the large sample of events from the 1995 (``-0.75T'') run with a 
single antiproton within the experimental acceptance, there are some 
events with two identified
antiprotons in the same event.  These two antiproton events give
insight into the possible correlated production of antimatter.  Since
the number of individual nucleon-nucleon collisions in each Au + Pb
collision is large, if the sample of central events are similar in
nature, the production of one antiproton should have very little
relation to the production of a second antiproton. 

In the 1995 data set there are approximately 43,000 antiprotons with rapidity less than
2.2, which were considered for this study.  After corrections for background 
contributions, we find there are 3.8 events with two antiprotons.


If we assume that the production of antiprotons is uncorrelated, we
can calculate the number of two antiproton events expected.  
One can think of the nucleus-nucleus
collision as many ($n$) nucleon-nucleon collisions each with a
probability ($p$) of producing an antiproton.  Since the probability ($p$) is
small and the number of collisions ($n$) is large, we
calculate the probability of producing two antiprotons in the same
event using Poisson statistics.  
The probability of producing one antiproton is:
\begin{equation}
\rm{Prob(1)} = Rate_{Singles} = n \times p
\end{equation}
The probability of producing two antiprotons is:
\begin{equation}
\rm{Prob(2)} = {{\rm{Rate_{singles}}^2} \over {2}}
\end{equation}
Since we have measured the rate of single antiprotons into our
detector $\rm{Rate_{Singles}}$, we calculate the expected number of two
antiproton events at 1.8.  The 90\% confidence level upper limit on
this number is five, which includes the experimentally measured value. 

Given the agreement with the assumption of uncorrelated production,
there are limits we can set on the possible correlated production of
antimatter.  We postulate that there are two distinct classes of
events within the 10\% central Au+Pb sample:  One class of purely
hadronic reactions and one class with the formation of the quark-gluon
plasma (QGP).

\begin{figure}
\vspace{4cm}
\centerline{\hbox{\psfig{figure=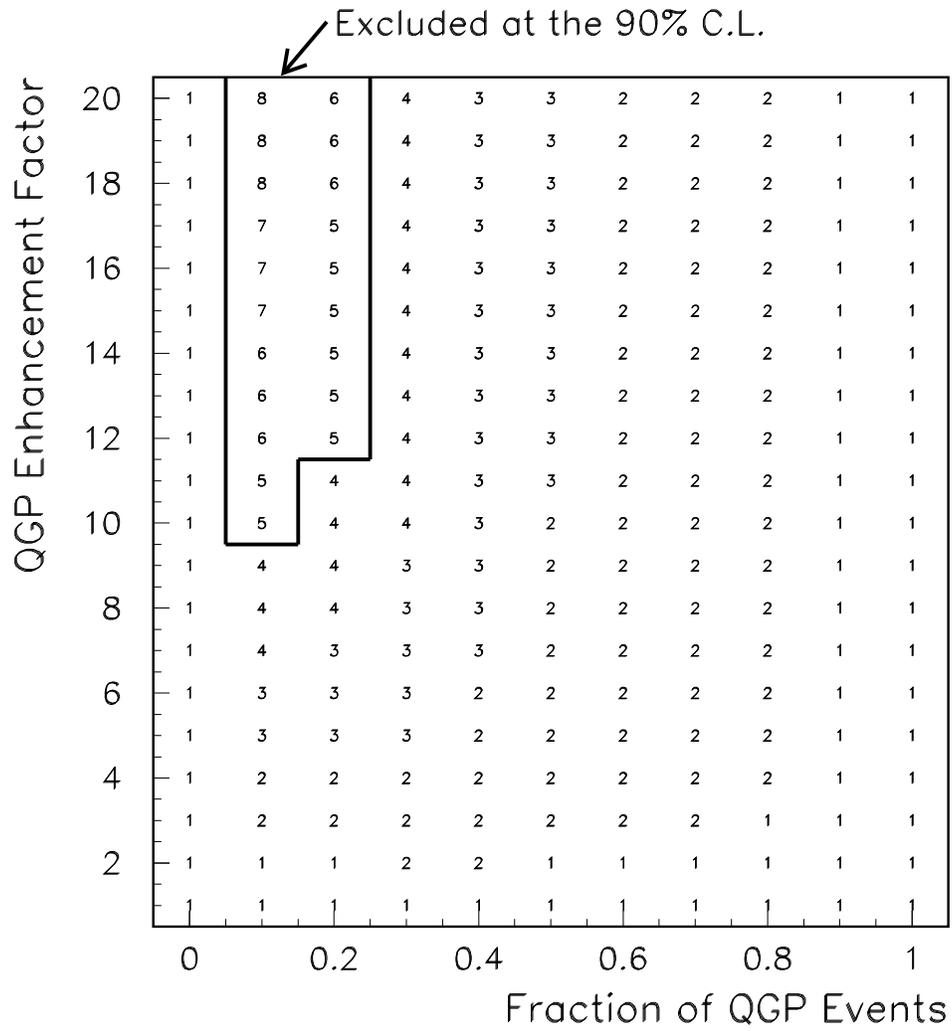,height=13.5cm}}}
\vspace{1cm}
\caption{The number of expected double antiproton events as a function
of the fraction of total events which have a QGP formation and the
relative antimatter enhancement factor in these QGP events to normal
hadronic events.} 
\label{fig:fig_pbar_doubles}
\end{figure}

In Figure \ref{fig:fig_pbar_doubles} the predicted number of
two antiproton events as a function of the fraction of QGP events
$f_{QGP}$ and the antimatter enhancement factor $\epsilon$ is shown.   
The area in the
dark box is where the predicted number of two antiproton events is
greater than five and thus ruled out by the data at the 90\%
confidence level.  If the QGP enhancement factor is small, the two
antiproton yield is not changed significantly.  Also, if most of the
events are QGP, then regardless of the enhancement, there is no
predicted increase in the two antiproton yield.  However, if there is
a large enhancement ($\epsilon >10$) and the fraction of QGP events is
between 5\% to 25\%, the yield of two antiprotons is significantly
increased.  These specific scenarios are ruled out by this measurement
at the 90\% confidence level. 

\subsection{Antideuteron Search}

\begin{figure}
\vspace{4cm}
\centerline{\hbox{\psfig{figure=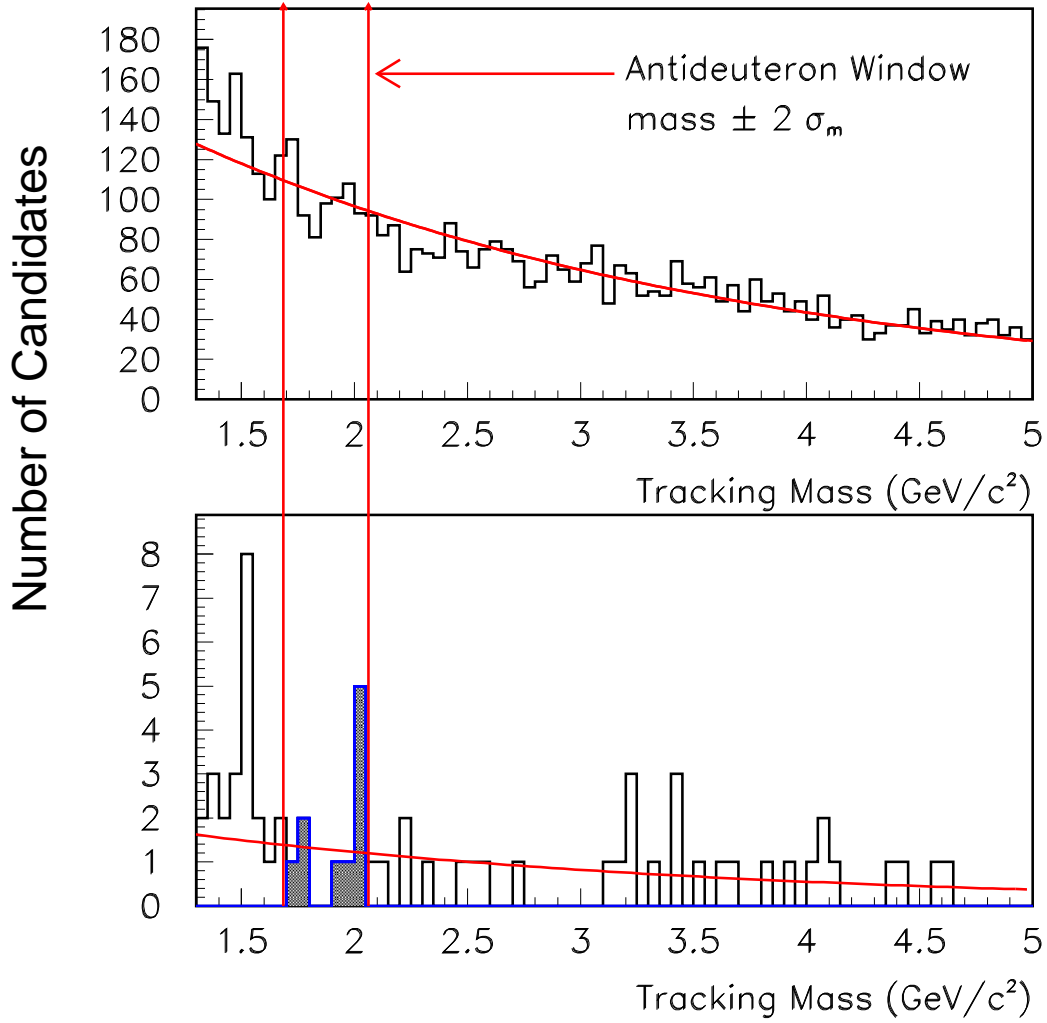,height=13.5cm}}}
\vspace{1cm}
\caption{The upper panel shows the antideuteron candidate tracking mass distribution with no requirement on
the calorimeter mass.  The bottom panels shown the same distribution after a cut requiring
the mass determined from the calorimeter cluster be larger than 1.6 GeV/$c^{2}$. The vertical
lines show the antideuteron mass window used in the analysis.} 
\label{fig:fig_dbar_mass}
\end{figure}

We have performed a search for antideuterons using the 1995 data set of
central Au + Pb interactions taken at the
``-0.75 T'' magnetic field setting optimized for the
acceptance of antideuterons.  After processing the data, any
tracked particle of charge negative one, rapidity $y < 2.4$, 
passing all track quality $\chi^{2}$ selections, and having a
reconstructed mass in the range $1.3~<~m~<~5$~$\rm{GeV/c^{2}}$
is considered a possible antinuclei candidate.  The mass
distribution of these candidates is shown in Figure \ref{fig:fig_dbar_mass}.  The
distribution is well fit by an exponential and has no significant
signal at the antideuteron mass $m = 1.874$ GeV/c$^{2}$.

The experiment is able to reduce this background through an
energy measurement using our full coverage hadronic calorimeter.
The calorimeter measures the deposited kinetic energy of
hadrons in addition to the annihilation energy for antibaryons.
The background processes expected to create high mass candidates
in the tracking reconstruction are the result of neutrons
which charge exchange in the vacuum exit window or air and
produce a forward going proton traversing the downstream spectrometer.
The protons have reasonable rapidity values, but reconstruct to
erroneously large rigidities resulting from the incorrect assumption
that the particle originated at the target.  These candidates 
should leave significantly less energy in the calorimeter than
expected if they are actually protons compared with real antideuterons
or heavier antinuclei.

Since these candidates are all assumed to be antimatter, the
reconstructed calorimeter mass must account for the annihilation
contribution.
In studies of antiproton showers from test beam data and from
the 1995 data, it was observed that only $\approx$ 84\% of the
annihilation energy was recorded in the calorimeter.  Thus, the calorimeter mass
formula is modified to reflect this loss:
\begin{equation}
\label{eqn:eqn_anti_mass}
\rm{mass} = {{E} \over {\gamma + 0.68}}.
\end{equation}

\begin{figure}
\vspace{4cm}
\centerline{\hbox{\psfig{figure=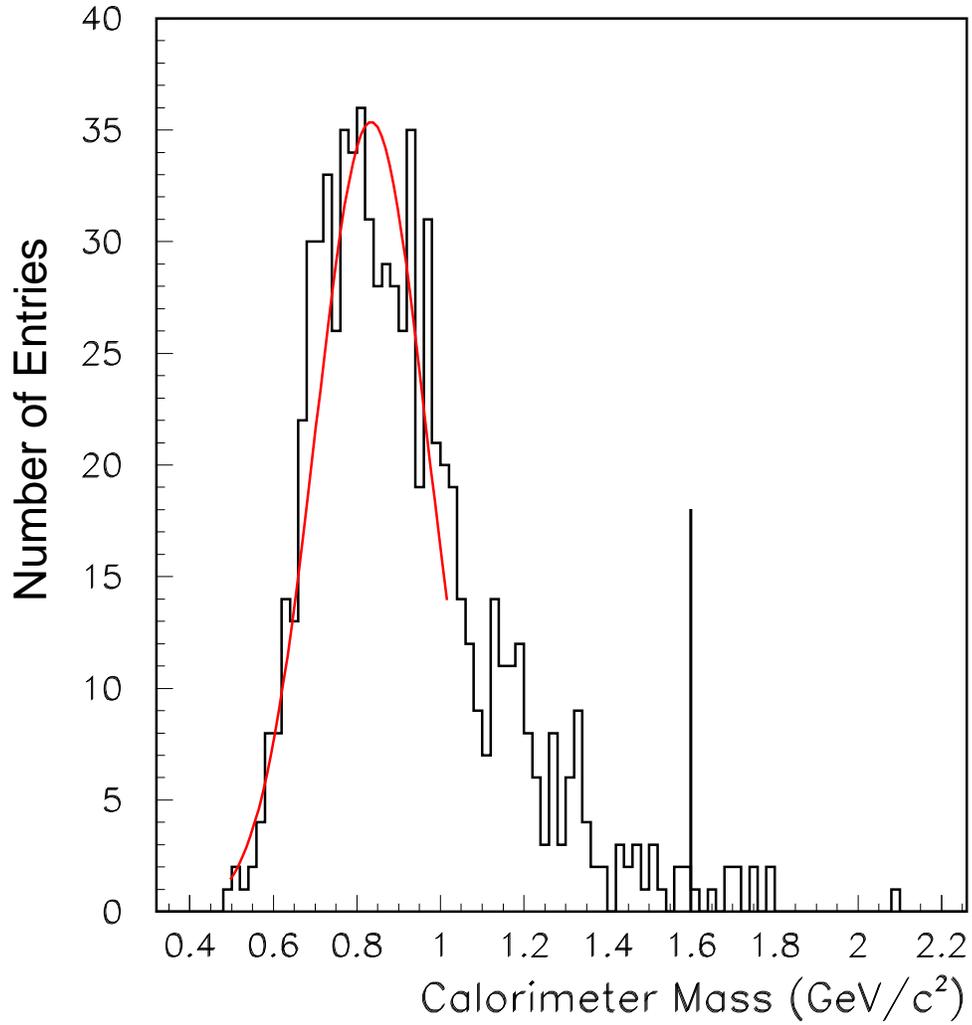,height=13.5cm}}}
\vspace{1cm}
\caption{Calorimeter reconstructed mass for antideuteron candidates
with tracking mass $1.687<m<2.061$~$\rm{GeV/c^{2}}$. The line shows the mass cut at 
1.6 GeV/c$^{2}$ applied in the analysis.} 
\label{fig:fig_dbar_camass}
\end{figure}

The tracking mass resolution is ${{\Delta m} \over {m}}\simeq5\%$, which
yields a $\sigma_{m}= 0.094$~$\rm{GeV/c^{2}}$ for antideuterons.  The
distribution of calorimeter masses for candidates whose tracking mass
is within $\pm 2\sigma_{m}$ of the antideuteron mass ($1.687 < m <
2.061~\rm{GeV/c^{2}}$) is shown in  Figure \ref{fig:fig_dbar_camass}.
One can see the peak mean value is less than $0.938$~$\rm{GeV/c^{2}}$.
Protons have a lower calorimeter mass ($<0.938$~$\rm{GeV/c^{2}}$)
when calculated using Equation \ref{eqn:eqn_anti_mass}  since they do
not deposit any energy beyond their kinetic energy (there is no
annihilation energy contribution).  The background candidates appear to
be protons  as expected from charge exchange background.  Most protons
should $\bf{\rm{not}}$ reconstruct such a large antimatter mass and
fire the late-energy trigger.  However, the calorimeter energy
response has a non-Gaussian high side tail.  These candidates are
protons which occupy the high side tail part of the energy response
distribution.  As can be seen in the plot, the calorimeter is a
powerful tool for rejecting this proton background. A cut is then
placed on calorimeter mass being greater  than 1.600~$\rm{GeV/c^{2}}$. 

If one assumes that all of the observed candidates are from charge
exchange background (really protons striking the calorimeter), then
the background shape can be fit.   The tracking mass distribution with
no cut on the calorimeter mass is fit to a simple exponential function
as shown in Figure \ref{fig:fig_dbar_mass}.  If the candidates are all
protons striking the calorimeter, the calorimeter mass distribution
should be the same regardless of the tracking mass.  Thus, one can use
the exponential function fit parameters  from the tracking mass
distribution with no calorimeter cuts to describe the tracking mass
distribution with a calorimeter mass cut. 

The tracking mass distribution is plotted in Figure \ref{fig:fig_dbar_mass} with a cut on the
calorimeter mass greater than 1.600~$\rm{GeV/c^{2}}$.  There are ten  candidates
within the $\pm 2 \sigma_{m}$ range of the antideuteron.  The
background fit distribution shown in Figure \ref{fig:fig_dbar_mass} is
renormalized to the total number of counts and plotted.   The
exponential fit seems a reasonable description of the distribution.
The total number of counts from the fit in the region of the
antideuteron (within $\pm 2 \sigma_{m}$) is 9.0.  Thus, there is no
significant signal above background for the antideuteron.  

One can then ask, how many real antideuterons would there have to be
to make a statistically significant peak above the background
distribution.  
There are nine predicted background events, and thus the Poisson
statistics 90\% confidence level upper limit is 14.2.  If more than
14.2 candidates were observed within the antideuteron mass range,
there is less than a 10\% chance that it is due to a statistical
fluctuation in the background events.  Thus, we set the 90\%
confidence level upper limit on antideuteron production at
$N_{Poisson}-N_{Background} = 5.2$. 

In order to translate this Poisson statistics limit into a total upper
limit on the production of antideuterons, the various acceptances and
efficiencies must be known.  
It is also possible using a specific
production model to set the 90\% confidence level upper limit on the
invariant multiplicity in a given region of momentum space.  In the 
discussion that follows, we will asuume a model in which the production
in $p_{T}$ and rapidity ($y$) can be factored as:
\begin{equation}
\frac{1}{2\pi p_{T}} \frac{dN}{dydp_{T}} = A_{0}e^{-2p_{T}/<p_{T}>}e^{-(y-y_{cm})^{2}/2{\sigma_{y}}^{2}}
\end{equation}
We have assumed a rapidity width $\sigma_{y} = 0.5$ and a mean transverse momentum
$<p_{T}> = 1.00$ GeV/c.  
Using this production model, we set a 90\% confidence level upper limit
on the production of antideuterons at $2.78 \times 10^{-7}$ per 10\% central 
Au + Pb interaction.

One can relate the limit over all phase space to the limit on the invariant
multiplicity ($A_{0}$) at midrapidity and $p_{T}=0$. 
\begin{equation}
A_{0} = {{N_{\rm{Total \space Limit}}} \over {(2\pi)^{3/2} \sigma_{y}
{{<p_{T}>^{2}} \over {4}}}}
\end{equation}
The upper limit on the invariant multiplicity at midrapidity $y=1.6$
and $p_{t}=0$ is $1.4 \times 10^{-7} \rm{GeV^{-2}c^{2}}$.  
We have tested the model dependency of these upper limits and
find that with extreme ranges of production models, one can vary 
the upper limits by approximately $\pm$50\%. 

\begin{figure}
\vspace{4cm}
\centerline{\hbox{\psfig{figure=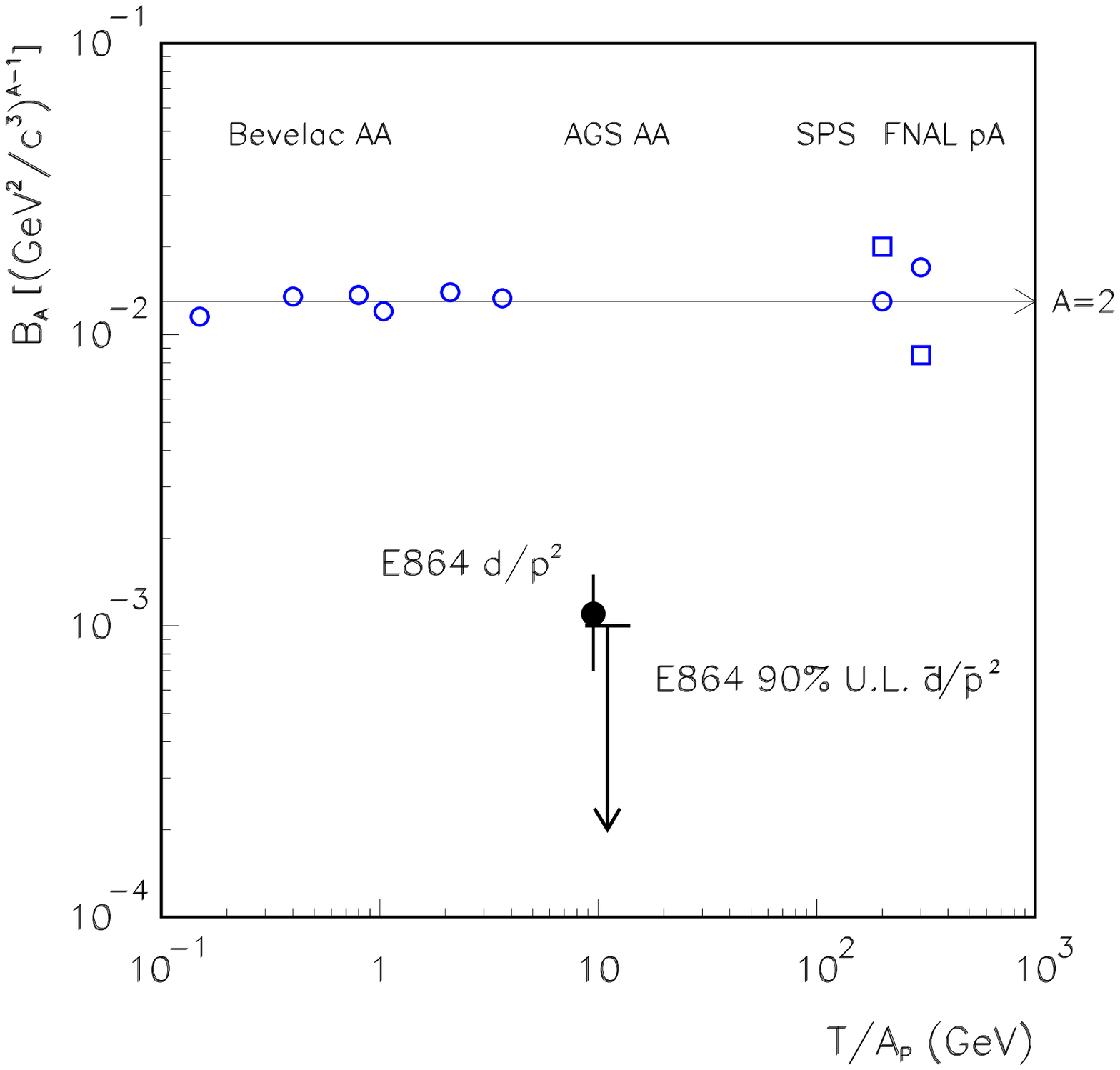,height=13.5cm}}}
\vspace{1cm}
\caption{Coalescence scale factors as a function of kinetic energy per
nucleon of the colliding beam. The upper limit on $\overline{B_{2}}$
from E864 is shown as an arrow. The E864 value for
deuterons $B_{2}$ is also shown. The simple coalescence level is
shown as a line with low energy AA and high energy pA results.} 
\label{fig:fig_dbar_coal}
\end{figure}

Using our antiproton measurements and these upper limits, we calculate 
the 90\% confidence level upper limits on the
coalescence scale factor $\overline{B_{2}}$ for antideuterons.  This
scale factor may be a function of where in momentum space the
measurement is made, thus we give the limit at midrapidity ($y=1.6$)
and $p_{t}=0$.  Our measured invariant multiplicity for antiprotons is
$1.16 \times 10^{-2}~\rm{GeV^{-2}c^{2}}$ (from the combined 10\% central
1995 and 1996 data).  The upper limit on the
invariant multiplicity for antideuterons is $1.41 \times
10^{-7}~\rm{GeV^{-2}c^{2}}$.  The upper limit on the
scale factor is:
\begin{equation}
\overline{B_{2}} = { { \left[{{1} \over {2\pi p_{t}}} {{d^{2}N} \over
{dydp_{t}}}(\overline{d}) \right]} \over { \left[{{1} \over {2\pi
p_{t}}} {{d^{2}N} \over {dydp_{t}}}(\overline{p}) \right]^{2}} } \leq
1.0 \times 10^{-3}~\rm{GeV^{2}c^{-2}}
\end{equation}
This upper limit is shown as an arrow in
Figure \ref{fig:fig_dbar_coal}, along with a comparison to coalescence scale
factors measured at Bevelac and SPS energies \cite{fig16_ref1} \cite{fig16_ref2} 
\cite{fig16_ref3}.

This scale factor is significantly below the global value of $1.2
\times 10^{-2}~\rm{GeV^{2}c^{-2}}$ predicted by the ``simple''
coalescence model.  However, since this prescription has failed to
describe systems where the collision volume is expected to be large
compared with the deuteron/antideuteron size \cite{e814_prc}, 
it is not surprising that it is in disagreement with the value obtained here.  

If the source distribution of antinucleons has a similar spatial
extent as the nucleon source, then the scale factor for deuterons
$B_{2}$ is expected to be the same as for antideuterons
$\overline{B_{2}}$.  Recently, E864 has presented
measurements of protons and deuterons around midrapidity
and low transverse momentum.   The scale factor from the
analysis of E864 light ion data \cite{Nigel_thesis} is also shown in
Figure \ref{fig:fig_dbar_coal}. 
\begin{equation}
B_{2} = 1.1\pm.4 \times 10^{-3}~\rm{GeV^{2}c^{-2}}
\end{equation}
The uncertainties are dominated by systematic errors in the deuteron
and proton invariant multiplicities.  This measured scale factor is at
the same level as the upper limit for the antideuteron scale factor.
We cannot determine whether $\overline{B_{2}}$ is significantly lower
than $B_{2}$.  Thus, it is impossible to comment on whether the rate
of antideuteron production is smaller due to preferential surface
emission of antimatter. 

If we consider the most probable value of the ratio
$(\overline{Y})/\overline{p} = 3.5$ for 10\%
central collisions, the primordial antiproton multiplicity at
midrapidity and $p_{t}=0$ should be a factor of $\sim3.3$ lower than
measured in E864.  In this picture, the 90\% confidence level upper
limit on the coalescence scale factor would be
\begin{equation}
\overline{B_{2}} \leq  1.1 \times 10^{-2}~\rm{GeV^{2}c^{-2}}
\end{equation}
This value is approximately at the level measured in p + A collisions,
in accord with the ``simple'' coalescence model, where the collision
volume is expected to be quite small.  Thus, if the
$\overline{Y}$ production is correctly calculated, the limit set
on antideuteron production is not very significant in the context of
coalescence models. 

There have been two previous measurements of antideuteron production in 
heavy ion collisions.  The first was from the E858 experiment which observed 
two antideuterons in minimum bias Si + Au collisions at 
14.6 A GeV/c\cite{e858_ref}.   They calculated a coalescence factor of 
approximately $\overline{B_{2}} \leq 1.0 \times 10^{-3}~\rm{GeV^{2}c^{-2}}$.  
While this value is consistent with our observation, it is difficult to 
make any direct comparison since the E858 value is for minimum bias 
collisions involving a much smaller projectile.  The second measurement 
is from experiment NA52 in Pb + Pb central collisions 
at 160 A GeV/c\cite{na52_ref}.  They observe a coalescence factor of 
approximately $\overline{B_{2}} \approx 5.0 \pm 3.0 \times 10^{-4}~\rm{GeV^{2}c^{-2}}$.  
They also find the factor for deuterons $B_{2}$ is the same within 
statistical and systematic errors.  While our upper limit on antideuterons 
is consistent with their value, our deuteron coalescence factor is 
somewhat higher.  This observation is not surprising due to larger source 
dimensions in the higher energy collisions studied by NA52. 

\section{Conclusions}

We have presented results from Experiment 864 for antiproton production and
antideuteron limits in Au+Pb collisions at 11.5 GeV/c per nucleon.  We
have measured invariant multiplicities for antiprotons above midrapidity
and at low transverse momentum as a function of collision geometry.  
These measurements are within systematic errors of
our previously reported results \cite{jlprl}, and, when compared with
the results from Experiment 878, may indicate a significant
contribution to the measured antiproton yield from the decay of
strange antibaryons.  


We have also studied correlated production of antimatter using 
events with more than one antiproton and a search for antideuterons.
For antideuterons we see no statistically significant signal.  
We set upper limits on the production at approximately 
$3 \times 10^{-7}$ per 10\% highest multiplicity Au+Pb interaction.

\begin{table}
\begin{center}
\begin{tabular}{|c|c|}
 rapidity & {$\frac{1}{2\pi p_{T}} \frac{dN}{dydp_{T}}$ at $p_{T}=0$ ($\times 10^{-2}$, in $\rm{GeV^{-2}c^{2}}$)} \\ \hline \hline
{ \hspace{1.5cm} $1.8<y<2.0$ \hspace{1.5cm} } &{ $1.00\pm0.05$ } \\
  $2.0<y<2.2$   &  $0.67\pm0.03$   \\
  $2.2<y<2.4$   &  $0.36\pm0.02$   \\
\end{tabular}   
\vspace{3.0mm}           
\caption{Antiproton invariant multiplicities at $p_{T}=0$ from the 1995 (-0.75T) 10\% 
central data. The errors listed are statistical only.} 
\label{tab:pbar_invar95}
\end{center}
\end{table}

\begin{table}
\begin{center}
\begin{tabular}{*{6}{|c}|}
 & \multicolumn{5}{c|}
{ {$\frac{1}{2\pi p_{T}} \frac{dN}{dydp_{T}} $ at $p_{T}=0$ ($\times 10^{-2}$, in $\rm{GeV^{-2}c^{2}}$)} } \\ \hline
 rapidity        & {100-70\%}     & {70-30\%}   & {30-10\%}  & {10\%}   & { min. bias } \\ \hline \hline
{ $1.4<y<1.5$ }  & { $0.12\pm0.02$ } & { $0.56\pm0.06$ } & { $0.87\pm0.08$ } & { $0.94\pm0.11$ } & { $0.48\pm0.02$ }\\
 $1.5<y<1.6$     &   $0.14\pm0.01$   &   $0.55\pm0.02$   &   $0.98\pm0.04$   &   $1.17\pm0.06$   & $0.51\pm0.01$ \\
 $1.6<y<1.7$     &   $0.13\pm0.01$   &   $0.53\pm0.02$   &   $0.90\pm0.03$   &   $1.15\pm0.05$   & $0.48\pm0.01$ \\
 $1.7<y<1.8$     &   $0.12\pm0.01$   &   $0.52\pm0.02$   &   $0.80\pm0.02$   &   $1.11\pm0.03$   & $0.46\pm0.01$ \\
 $1.8<y<1.9$     &   $0.11\pm0.01$   &   $0.44\pm0.02$   &   $0.75\pm0.02$   &   $1.00\pm0.04$   & $0.39\pm0.01$ \\
 $1.9<y<2.0$     &   $0.093\pm0.005$   & $0.38\pm0.02$   &   $0.68\pm0.02$   &   $0.88\pm0.03$   & $0.34\pm0.01$ \\
 $2.0<y<2.1$     & { $0.060\pm0.004$ } & $0.30\pm0.01$   &   $0.51\pm0.02$   &   $0.71\pm0.03$   & $0.27\pm0.01$ \\
\end{tabular}              
\vspace{3.0mm}
\caption{Antiproton invariant multiplicities at $p_{T}=0$ vs. centrality from the 1996 (-0.45T) minimum-bias
data.  The data are listed by multiplicity regions used in the analysis.
The invariant multiplicities for the full minimum-bias 
sample are also listed.  The errors listed are statistical only.} 
\label{tab:pbar_invar96}
\end{center}
\end{table}


\begin{references}

\bibitem[\ast]{}        Present address: Vanderbilt University, Nashville, Tennessee 37235 
\bibitem[\dag]{}        Present Address: Anderson Consulting, Hartford, CT
\bibitem[\ddag]{} 	Present address: Univ. of Denver, Denver CO 80208
\bibitem[\S]{}    	Deceased.
\bibitem[\|]{}     	Present address: McKinsey \& Co., New York, NY 10022
\bibitem[\P]{}          Present address: Department of Radiation Oncology, Medical College of Virginia, Richmond VA 23298
\bibitem[\ast\ast]{}	Present address: Columbia University, Nevis Laboratory, Irvington, NY 10533
\bibitem[\dag\dag]{}    Present address: University of Tennessee, Knoxville TN 37996
\bibitem[\ddag\ddag]{}  Present address: Institut de Physique Nucl\'{e}aire, 91406 Orsay Cedex, France
\bibitem[\S\S]{} 	Present Address: Institute for Defense Analysis, Alexandria VA 22311
\bibitem[\|\|]{} 	Present Address: MIT Lincoln Laboratory, Lexington MA 02420-9185

\bibitem{qgprefs}P. Koch, B. M\"{u}ller, H. St\"{o}cker and W. Greiner, 
Modern Physics Letters A \textbf{8} 737 (1988);
J. Ellis, U. Heinz and Henry Kowalski, Phys. Lett. B \textbf{233} 223 (1989);
U. Heinz, P. R. Subramanian, H. St\"{o}cker and W. Greiner, Journal of 
Physics G: Nuclear Physics \textbf{12}, 1237 (1986).

\bibitem{mean}V. Koch, G. E. Brown and C. M. Ko, Phys. Lett. B \textbf{265} 
29 (1991); J. Schaffner, I. N. Mishustin, L. M. Satarov, H. St\"{o}cker and W. Greiner, 
Z. Phys. A \textbf{341} 47 (1991); P. Koch and C. B. Dover, Phys. Rev. C \textbf{40} (1989) 145;
S. Gavin, M. Gyulassy, M. Pl\"{u}mer and R. Venugopalan, Phys. Lett. B 
\textbf{234} 175 (1990).

\bibitem{arc_pap}S. H. Kahana, Y. Pang, T. Schlagel and C. B. Dover, 
Phys. Rev. C \textbf{47} 1356 (1993).

\bibitem{Nagle_prl}J.L. Nagle, B.S. Kumar, M.J. Bennett, S.D. Coe, G.E. Diebold, and J.K. Pope, Phys. Rev. Lett. \textbf{73} 2417 (1994).

\bibitem{e864_nimpap}T. Armstrong \emph{et al.}, to be submitted to NIM.

\bibitem{calo_nim}T. Armstrong \emph{et al.}, Nucl. Instr. Meth. A \textbf{406} (1998) 227.

\bibitem{let_nimpap}J. Hill \emph{et al.}, accepted for publication in NIM.

\bibitem{beam_nim}P. Haridas, G. Van Buren, J. Tomasi, M.S.Z. Rabin, K. Barish, and R.D. Majka, Nucl. Instr. Meth. A \textbf{385} (1997) 412.

\bibitem{GEANT} GEANT Detector Description and Simulation Tool, CERN Program
Library Long Writeup W5013.

\bibitem{nagle_thesis}J. Nagle, Ph.D. Thesis, Yale University (1997).

\bibitem{jlprl} T. Armstrong \emph{et al.}, Phys. Rev. Lett. \textbf{79} 3351 (1997);
J. Lajoie, Ph.D. Thesis, Yale University (1996).

\bibitem{e878_prl}D. Beavis \emph{et al.}, Phys. Rev. Lett. \textbf{75} 
3633 (1995).

\bibitem{mike_prc}M. Bennett \emph{et al.}, Phys. Rev. C \textbf{56} 1521 (1997).

\bibitem{costales}J. Costales, Ph.D. Thesis, Massachusetts Institute of Technology (1990)

\bibitem{e878_mult_nim} D. Beavis \emph{et al.}, Nucl. Instrum. Meth. A \textbf{357}, 283 (1995).

\bibitem{e886} G.E. Diebold \emph{et al.}, Phys. Rev. C \textbf{48} 2984 (1993).

\bibitem{e866_pbar}The E866 Collaboration, H. Sako \emph{et al.}, in 
\emph{Heavy Ion Physics at the AGS (HIPAGS 96)}, C. A. Pruneau, G. Welke, 
R. Bellweid, S. J. Bennett, 
J. R. Hall and W. K. Wilson Eds., WSU-NP-96-16, Wayne State University, 
Dec. 1996.

\bibitem{RQMD} H. Sorge, Phys. Rev. C \textbf{52}
3291 (1995); M. Gonin \emph{et al.}, Phys. Rev. C \textbf{51} 310 (1995).

\bibitem{Yeudong_Wu}The E859 Collaboration, Y. Wu \emph{et al.}, in 
\emph{HIPAGS 96}, see ref \cite{e866_pbar}.

\bibitem{pp_ref}V. Blobel \emph{et al.}, Nuclear Physics B \textbf{69} 
454 (1974); A. M. Rossi \emph{et al.}, Nuclear Physics B \textbf{84} 
269 (1975).

\bibitem{str_enh_refs}P. Koch, B. Muller, J. Rafelski, Phys. Rep. \textbf{142}, 167 (1986).

\bibitem{hgas_refs}P. Braun-Munzinger, J. Stachel, J. P. Wessels and N. Xu, 
Phys. Lett. B \textbf{344} 43 (1995).

\bibitem{gerd} G. J. Wang, G. Welke, R. Bellwied, and C. Pruneau, nucl-th/9807036


\bibitem{fig16_ref1}S. Nagamiya \emph{et al.}, Phys. Rev. C \textbf{24}, 971 
(1981); P. Lemaire, Phys. Rev. Lett. \textbf{37}, 667 (1976); 
M Anikina, JINR Report No. 1-84-216, Dubna (1984).

\bibitem{fig16_ref2}W. Bozzoli, Nucl. Phys. B \textbf{144}, 317 (1978).

\bibitem{fig16_ref3}J.W. Cronin \emph{et al.}, Phys. Rev. D \textbf{11}, 3105 (1975).

\bibitem{e814_prc} J. Barrette \emph{et al.}, Phys. Rev. C \textbf{50}, 1077 (1994);  
J.L. Nagle, B.S. Kumar, M.J. Bennett, S.D. Coe, G.E. Diebold, and J.K. Pope, Phys. Rev. Lett. \textbf{73}, 1219 (1994)

\bibitem{Nigel_thesis}N. K. George, Ph. D. Thesis, Yale University (1998)

\bibitem{e858_ref}A. Aoki \emph{et al.}, Phys. Rev. Lett. \textbf{69}, 2345 (1992).

\bibitem{na52_ref}G. Ambrosini \emph{et al.}, Nucl. Phys. A \textbf{638}, 411c (1998).

\end{references}
\end{document}